\documentclass[a4paper,11pt]{article}

\usepackage[utf8]{inputenc}

\usepackage{amsmath,amsfonts,amsthm,amssymb,dsfont}
\usepackage{graphicx,wrapfig,lipsum}
\usepackage[section]{placeins}
\usepackage{slashed}

\usepackage[numbers,sort&compress]{natbib}

\usepackage{tikz}
\usetikzlibrary{shapes.geometric,decorations.markings}
\usepackage{subfigure}

\usepackage[hidelinks]{hyperref}

\hypersetup{
    colorlinks=true,
    linktocpage=true,
    linkcolor=blue,
    urlcolor=blue,
    citecolor=blue,
}

\usepackage{geometry}
\usepackage{setspace}

\hypersetup{linkcolor=blue}
\phantomsection 
\addtocontents{toc}{\protect\addvspace{4.5pt}}

\numberwithin{equation}{section}
\newcommand{\Comment}[1]{{}}

\newcommand{\be}{\begin{equation}}
\newcommand{\bea}{\begin{eqnarray}}

\newcommand{\ee}{\end{equation}}
\newcommand{\eea}{\end{eqnarray}}

\def\d{\partial}

\def\Tr{{\rm Tr\, }}

\DeclareMathAlphabet{\mathpzc}{OT1}{pzc}{m}{it}

\global\long\def\dd{\mathrm{d}}%
\global\long\def\not#1{\slashed{#1}}%

\global\long\def\ri{\mathrm{i}}%

\newcommand{\Sp}{\text{S}}
\newcommand{\RR}{\text{R}}
\newcommand{\FF}{\text{F}}

\begin{document}

\begin{titlepage}

\begin{flushright}
    \par
\end{flushright}

\vskip 0.5cm

\begin{center}
\begin{spacing}{2.4}

\textbf{\huge Moduli space of $\mathcal{N}=4$ super Yang-Mills from AdS/CFT}\\

\end{spacing}

\vskip 5mm

\vskip 1cm

\large {\bf Andr\'{e}s Anabal\'{o}n}$^{~a,b}$     
    \footnote{anabalo@gmail.com}, 
\large {\bf Horatiu Nastase}$^{~b}$   
    \footnote{horatiu.nastase@unesp.br},\vspace{0.4cm}
\large {\bf Carlos Nunez}$^{~c}$
    \footnote{c.nunez@swansea.ac.uk}, \\
\large {\bf Marcelo Oyarzo }$^{~a,d}$\footnote{moyarzoca1@gmail.com}  and
\large {\bf Ricardo Stuardo }$^{e}$\footnote{ricardostuardotroncoso@gmail.com}

\vskip .8cm 

$^{(a)}${\textit{Departamento de F\'isica, Universidad de Concepci\'on,\\ Casilla, 160-C, Concepci\'on, Chile.}}\\ \vskip .3cm
$^{(b)}${\textit{Instituto de Física Teórica, UNESP-Universidade Estadual Paulista\\ R. Dr. Bento T. Ferraz 271, Bl. II, Sao Paulo 01140-070, SP, Brazil.}}\\ \vskip .3cm
$^{(c)}${\textit{Centre for Quantum Fields and Gravity, Department of Physics, Swansea University,\\ Swansea SA2 8PP, United Kingdom}}\\ \vskip .3cm
$^{(d)}${\textit{Instituto Galego de Física de Altas Enerxías (IGFAE), Universidade de Santiago de Compostela,
E-15782 Santiago de Compostela, Spain}}\\ \vskip .3cm
$^{(e)}${\textit{Departamento de Física, Universidad de Oviedo,
Avda. Federico García Lorca 18, 33007 Oviedo, Spain \\ \vspace{0.1cm}
and\\ \vspace{0.1cm}
Instituto Universitario de Ciencias y Tecnologías Espaciales de Asturias (ICTEA), Calle de la Independencia 13, 33004 Oviedo, Spain
}}
\end{center}

\vskip .5cm 
\begin{abstract}
We study $\mathcal{N}=4$ super Yang–Mills theory compactified on an $\Sp^1$ at zero temperature, with VEVs for two scalar bilinears and three independent current sources. We show that type IIB supergravity provides a complete holographic description of this setup, admitting both supersymmetric and non-supersymmetric AdS-soliton solutions that are asymptotically AdS$_5$ and smooth in the IR. The current sources correspond in 2+1 dimensions to Q-ball charge densities for $\text{U(1)}^3 \subset \text{SO(6)}_\RR$, and are geometrically realized as twists along three angular directions of the $\Sp^5$. We demonstrate that the bulk dynamics encodes the full vacuum structure of the dual field theory and explicitly reconstruct the supersymmetric moduli space.
%
%
%
%
%
%
%
\end{abstract}

\vfill{}
\vspace{1.5cm}
\end{titlepage}

\setcounter{footnote}{0}
\tableofcontents

\section{Introduction}

Strongly coupled gauge theories are notoriously difficult to analyze from first principles. The AdS/CFT correspondence \cite{Maldacena:1997re, Witten:1998qj} circumvents this difficulty by recasting questions in the gauge theory in terms of a tractable dual supergravity description. If the geometry is everywhere regular, it means that the large-$N$, strongly coupled behavior of the QFT is fully captured by the supergravity theory. However, for example, the standard holographic description of the Coulomb branch of gauge theories is singular in the infrared \cite{Cvetic:1999xx, Cvetic:2000zu, Freedman:1999gk}.

It was recently found that, by compactifying the gauge theory on an $\Sp^1$ and giving a VEV to the R-symmetry current, the infrared geometry can end smoothly, yielding a supersymmetric ground state \cite{Anabalon:2021tua} of the AdS-soliton type \cite{Horowitz:1998ha}. These solutions describe the confining regime of the field theory, and the validity of the supergravity approximation has been explored in a large number of papers \cite{Nunez:2023xgl,Fatemiabhari:2024aua,Chatzis:2024top,Chatzis:2024kdu,Giliberti:2024eii,Castellani:2024ial,Macpherson:2024qfi,Barbosa:2024smw,Kumar:2024pcz,Nunez:2023nnl,Fatemiabhari:2024lct,Jokela:2025cyz,Nunez:2025gxq,Chatzis:2025dnu,Macpherson:2025pqi,Nunez:2025ppd,Chatzis:2025hek,Fatemiabhari:2025usn,Fatemiabhari:2026goj}.  

The existence of two solutions, singular in the infrared and exhibiting qualitatively different behavior while sharing the same boundary conditions, was noticed in the early holographic explorations of the Coulomb branch \cite{Freedman:1999gk}. In \cite{Anabalon:2024che}, IR-regular geometries with a single supergravity scalar were constructed, and it was shown that the phase boundary between the different Coulomb branch behaviors corresponds to the point at which the scalar VEV vanishes.
Imposing UV boundary conditions fixes the asymptotic data, but regularity in the IR further constrains the solutions, determining the VEVs as functions of the sources and the energy scale. This structure naturally gives rise to a moduli space at each energy, consisting of all solutions compatible with a given set of sources. The locus where the scalar VEV vanishes signals a phase transition in the theory, a phenomenon recently shown to take place for supergravity black holes as well \cite{Anabalon:2024lgp,Anabalon:2025sok}.
In particular, the requirement of IR regularity dynamically selects admissible vacua, thereby promoting what would naively be arbitrary integration constants into physically meaningful order parameters of the dual theory.

In this paper we begin by constructing a soliton solution in 
the STU model of type IIB supergravity compactified on the $\Sp^5$, 
with two independent scalars and three Abelian vectors. The solution has nontrivial supergravity scalars and  vectors. It generalizes all  previous constructions of AdS solitons within this model. This construction provides the first fully regular SUSY multi-charge solitonic completion of the Coulomb branch within the STU truncation, offering a controlled holographic laboratory to study confinement, symmetry breaking, and vacuum selection in strongly coupled gauge theories. The solution is found by means of double Wick rotation (together with a non-trivial diffeomorphism) and a reparametrization of the black holes of \cite{Behrndt:1998jd}, in such a way that they admit a massless limit. The massless sector contains the supersymmetric solutions we are interested in.
Unlike the previously known single-scalar constructions, our solution allows for the simultaneous backreaction of multiple scalar fields and independent ``charge'' densities, leading to a qualitatively richer phase structure and a non-trivial multi-parameter moduli space of supersymmetric vacua.

The uplift to ten-dimensional type IIB supergravity warps the 
five dimensional spacetime, and squashes and twists the $\Sp^5$. This allows an interpretation in the dual ${\cal N}=4$ SYM and its
$\Sp^1$ reduction to 2+1 dimensions, in terms of deformations by scalars 
in the ${\cal O}_{20'}$ multiplet and Q-ball charge densities.

The outline of this paper is as follows. In Section 2 we describe
the five-dimensional solution.
We present the Lagrangian and its associated solutions. Then we compute 
the VEVs and sources by means of an asymptotic analysis. 
We characterize the solution space and show that there are at most five solutions for each choice of boundary conditions. We then describe how to recover the previously known 
solutions. Finally we compute the moduli space of supersymmetric 
solutions for our new general soliton and we carry out the 
calculation of the Killing spinors.
In Section 3 we uplift the solution to ten dimensional type IIB supergravity. In 
Section 4 we interpret our new solution from the dual viewpoint of four-dimensional
${\cal N}=4$ SYM and its $\Sp^1$ twisted compactification to 2+1 dimensions. In 
Section 5 we present  conclusions and future directions.

\section{Five-dimensional Perspective}

In this section, we present a new family of solutions to the gauged STU model, which we subsequently uplift to type IIB supergravity. In both the five-dimensional and ten-dimensional descriptions, we have explicitly verified that all equations of motion are satisfied. We perform a detailed asymptotic analysis of the five-dimensional solutions and establish their interpretation as deformations of ${\cal N}=4$ SYM, regarded as the UV fixed point of the theory. We systematically explore the space of solutions, placing particular emphasis on the supersymmetry-preserving sector, for which we derive the explicit form of the preserved Killing spinors. For specific regions of parameter space, we demonstrate how our new solutions interpolate to previously known backgrounds.

\subsection{The model}
We shall study a truncation of type IIB supergravity compactified over a deformed $\Sp^5$ with action 
\begin{equation}
S_0=\frac{1}{2\kappa}\int\sqrt{-g}\left(  R-\frac{\left(
\partial\Phi_{1}\right)  ^{2}}{2}-\frac{\left(\partial\Phi_{2}\right)  ^{2}}{2}+\sum_{i=1}^{3}4L^{-2}X_i^{-1}-\frac{1}{4}X_{i}^{-2}(F^{i})^{2}+\frac{1}{4}\epsilon^{\mu \nu \rho \sigma \lambda}A^1_{\mu}F^2_{\nu \rho}F^3_{\sigma \lambda} \right)  \dd^{5}x,
\label{LSTU}%
\end{equation}
where $F^{i}$ are two forms, related with gauge fields in the standard
way $F^{i}=dA^{i}$, $X_{i}=e^{-\frac{1}{2}\vec{a}_{i}\cdot
\vec{\Phi}}$, $\vec{\Phi}=\left(  \Phi_{1},\Phi_{2}\right)  $ and%
\begin{equation}
\vec{a}_{1}=\left(  \frac{2}{\sqrt{6}},\sqrt{2}\right),\qquad\vec{a}_{2}=\left(  \frac{2}{\sqrt{6}},-\sqrt{2}\right),\qquad\vec{a}_{3}=\left(  -\frac{4}{\sqrt{6}},0\right).
\end{equation}
This is the gauged STU model in $D=5$. Let us  present the new family of solutions.

\subsection{New AdS Solitons}
We study a new family of solutions, depending on the parameters $(q_1,q_2,q_3, M, {\mathtt{q} })$. We activate the  two scalars $(\Phi_1,\Phi_2)$ and the three gauge fields $A^i$, exciting all possible fields in the STU model.
\subsubsection{Three Charge Solutions}
We  find the following solution of the gauged STU model 
in $D=5$. The metric tensor is given by
are
\begin{equation}\label{thebackground}
\dd s^{2} =\frac{f(r)}{H(r)^{2/3}}\dd\varphi^{2}+\frac{H(r)^{1/3}}{f(r)}\dd r^{2}+\frac{H(r)^{1/3}}{L^{2}}r^{2}(-\dd t^{2}+\dd x^{2}+\dd y^{2}).
\end{equation}
The scalars and the vectors are
\begin{equation}\label{A=000020Lambda}
\Phi_{1} =\frac{\sqrt{6}}{6}\log\left(\frac{H_{1}(r)H_{2}(r)}{H_{3}(r)^{2}}\right)\,,\quad
\Phi_{2}=-\frac{1}{\sqrt{2}}\log\left(\frac{H_{2}(r)}{H_{1}(r)}\right)\,, \quad
A^{i} =\left(\frac{Q_{i}}{r^{2}H_{i}(r)}- {\mu_i} \right)\dd\varphi\,.
\end{equation}
The metric function and the harmonic functions are
\begin{align}
f(r) & =\frac{r^{2}}{L^{2}}H(r)-\frac{M}{r^{2}}-\frac{\mathtt{q}}{r^{4}}\,,\qquad 
H(r)=H_{1}(r)H_{2}(r)H_{3}(r)\,,\qquad 
H_{i}(r)=1+\frac{q_{i}^{2}}{r^{2}}\,.
\end{align}
This configuration is a solution of the equations of motion provided
    \begin{equation}\label{eq:ParametersConstraint}
        Q_i^2 = -M q_i^2 + {\tt q}.
    \end{equation}

Compared to \cite{Cvetic:1999xp}, one novelty of this set of coordinates is that ${\tt q}\neq 0$, 
which has the property that the massless limit, $M=0$, yields a 
metric that might smoothly close at the origin, provided there is 
a $r_0$ where $f(r_0)=0$. This limit yields $Q_1^2=Q_2^2=Q_3^2$.

The regularity of the gauge fields at $r=r_{0}$, where by 
definition $f(r_{0})=0$, fixes the constant $\mu_i$ 
in (\ref{A=000020Lambda}) to be
\begin{equation}\label{eq:RegularityGaugeFields}
\mu_i =\frac{Q_i}{r_{0}^{2}H_{i}(r_{0})}\,.
\end{equation}
When  an $r_{0}$ exists, the configuration is regular everywhere. 
The solution allows for a good $M\to0$ limit, which yields new 
supersymmetric configurations, as we show further below. Now, let us focus on the asymptotic behavior of the family of backgrounds and relate this via holographic renormalisation \cite{Skenderis:2002wp, Papadimitriou:2004ap} to deformations of ${\cal N}=4 $ SYM.


\subsection{The Asymptotic Analysis}\label{sec:asymptotic}

To put the solution in standard asymptotically AdS form, 
we  require that, at least,
\begin{equation}
H(r)^{1/3} \frac{r^2}{L^2}=\frac{\rho^2}{L^2} + O(\rho^{-3})\, .
\end{equation}
The fall-off is fixed to gauge away all the possible 
contributions to the energy momentum tensor from the three-dimensional Minkowski part of the metric. We find that the 
asymptotic change of coordinate that implements this is
\begin{equation}
r = \rho 
-\frac{q_1^2 + q_2^2 + q_3^2}{6\rho}
+  (q_2 + q_1 - q_3)(q_1 + q_2 + q_3)(q_1 + q_3 - q_2)(-q_2 - q_3 + q_1) \frac{1}{24\rho^3}
+ O(\rho^{-5}).
\end{equation}
This yields 
\begin{equation}
g_{r r} \left(\frac{dr}{d\rho}\right)^2= g_{\rho \rho}=
\frac{L^2}{\rho^2} + \frac{1}{9}\frac{L^2(2q_1^2 q_2^2 + 2q_1^2 q_3^2 + 2q_2^2 q_3^2 - 2q_1^4 - 2q_2^4 - 2q_3^4 + 9ML^2)}{\rho^6} + O(\rho^{-8})\,,
\end{equation}
\begin{equation}
-g_{tt}=g_{yy}=g_{xx}= \frac{\rho^2}{L^2} + O(\rho^{-4}) \,,
\end{equation}
\begin{align}
g_{\varphi \varphi} &= \frac{\rho^2}{L^2}-\frac{M}{\rho^2}+O(\rho^{-4}) \,,
\end{align}
and for the gauge fields
\be
A_i=\left(\frac{Q_i}{\rho^2}-\mu_i  
+ \frac{Q_i(q_1^2 + q_2^2 + q_3^2 -3 q_i^2)}{3 \rho^4} + O(\rho^{-6})
\right)\dd\varphi.\label{Aiexp}
\ee
We plug this asymptotic expansion in the scalar fields. We get
\begin{align}
\Phi_1&=\frac{1}{\sqrt{6}}\frac{q_1^2+q_2^2-2q_3^2}{\rho^2}+O(\rho^{-4}) \, , \nonumber\\
\Phi_2&=\frac{1}{\sqrt{2}}\frac{q_1^2-q_2^2}{\rho^2}+O(\rho^{-4}) \, .\label{scalarexp}
\end{align}
These scalar fields saturate the Breitenlohner–Freedman bound, 
and therefore the source term is associated with the logarithmic 
mode, that is absent here. 

\subsection{Holographic Renormalization}

The analysis of the precise form of the VEVs, sources and the dual energy momentum tensor can be found in \cite{Anabalon:2024che}, which closely follow the standard references \cite{Balasubramanian:1999re, Bianchi:2001kw,Skenderis:2002wp}. Here we provide the formulae for further reference and facilitate the verification of our results. The action plus counterterms is
\begin{equation}
S = S_0 + \frac{1}{\kappa} \int_{M^3 \times S^1} K\sqrt{-h}\,  \dd^4x + \frac{1}{2\kappa} \int_{M^3 \times S^1} \sqrt{-h} \left( -\frac{6}{L} + \frac{1}{2L} \left( \frac{1}{\ln(\rho/\rho_0)} - 2 \right) (\Phi_1^2+\Phi_2^2) \right) \dd^4x \,, 
\end{equation}
where $S_0$ is the bulk action, $g_{\mu\nu} = h_{\mu\nu} + N_\mu N_\nu$, and $N_\mu$ is the outward pointing normal to the boundary and $K_{\mu\nu} = \frac{1}{2}\nabla_\mu N_\nu + \frac{1}{2}\nabla_\nu N_\mu$ is the extrinsic curvature. The boundary geometry is that of a three dimensional Minkowski spacetime times a circle,

\begin{equation}
\dd s^2 = \gamma_{ab} \dd x^a \dd x^b = -\dd t^2 + \dd y^2 + \dd x^2 + \dd \varphi^2 \,,\label{eq-2.15} 
\end{equation}
which is the background spacetime for the quantum field theory. The scalar fields have the general asymptotic expansion

\begin{equation}
\Phi_i = J_{\Phi_i} \frac{\ln(\rho/\rho_0)}{\rho^2} + \frac{\Phi_{0i}}{\rho^2} + O\left( \frac{\ln(\rho/\rho_0)}{\rho^4} \right) \,,
\end{equation}
with the on-shell variation

\begin{equation}
\frac{\delta S}{\delta J_{\Phi_i}} = \frac{1}{2\kappa L^5} \Phi_{0i}\, . 
\end{equation}
The finite scalar source is related to the gravity one by the relation $J_{\Phi_i}=L^4 J^{finite}_{\Phi_i}$. Therefore, we obtain the following VEVs, which are order $N^2$ in the QFT
\begin{align}
\langle \mathcal{O}_1 \rangle&= \frac{q_1^2+q_2^2-2q_3^2}{2\kappa \sqrt{6} L} \, , \\
\langle \mathcal{O}_2 \rangle&=\frac{ q_1^2-q_2^2}{2\kappa \sqrt{2} L}\, .\label{scalarexp2}
\end{align}
The vacuum expectation value of the energy momentum tensor of the dual field theory is
\begin{align}
\langle T_{ab} \rangle &= \frac{-2}{\sqrt{-\gamma}} \frac{\delta S}{\delta \gamma^{ab}} \nonumber
\\
&= \lim_{\rho \to \infty} \frac{\rho^2}{L^2} \frac{-2}{\sqrt{-h}} \frac{\delta S}{\delta h^{ab}} 
\nonumber\\
&= \lim_{\rho \to \infty} \frac{\rho^2}{L^2 \kappa} \left( h_{ab} K - K_{ab} - \frac{3}{L} h_{ab} - \frac{1}{2L} h_{ab} (\Phi_1^2+\Phi_2^2) \right) \,.
\end{align}

The scalars contribute to the dual 
energy-momentum tensor in a conformal invariant form and the 
results of \cite{Anabalon:2014fla} apply. The dual energy-momentum tensor is
\begin{equation}
\langle T_{tt} \rangle = -\frac{M}{2L^3 \kappa}, \qquad
\langle T_{xx} \rangle = \langle T_{yy} \rangle = \frac{M}{2L^3 \kappa}, \qquad
\langle T_{\varphi\varphi} \rangle = -\frac{3M}{2L^3 \kappa}.\label{tmunu}
\end{equation}

Let us now carefully study the space of solutions, parameterizing different quantities  in terms of the boundary values of fields.
\subsection{The Solution Space}
From the gravity point of view, to construct the space of 
solutions is equivalent to parameterize the observables in terms of the 
boundary conditions. The main observable here is the energy, 
which is controlled by the parameter $M$ in the metric, see eq.(\ref{tmunu}). On the other hand, the 
boundary conditions are the asymptotic values of the gauge fields 
$\mu_i$ and $\Delta$, the period of the coordinate  $\varphi \in [0,\Delta]$. This period 
is defined by the regularity condition (the space closes smoothly at $r=r_0$ where $f(r_0)=0$)
\begin{equation}
\Delta=\frac{4 \pi H(r_0)^{1/2}}{\lvert f'(r_0)\rvert} \, .\label{periododefi}
\end{equation}
The absolute value  in eq.(\ref{periododefi}) is important. Indeed,
\begin{eqnarray}
& & f'(r_0)=\frac{2}{r_0^{3} L^{2}}\left(h_{2}h_{3}+h_{1}h_{3}+h_{1}h_{2}-M L^{2}\right)\;,\label{fp-hi}
\end{eqnarray}
where we define
\begin{eqnarray}
h_{i} \equiv H_{i}(r_0) r_0^2\; .
\end{eqnarray}
The quantities $h_{i}$ 
have the advantage of being well defined at $r_0=0$ and $h_i\in 
\mathbb{R}^{+}$. Therefore, $f'(r_0)$ can, in principle, be 
negative when $M$ is large enough (although we shall see this is 
not the case). Hence, we use the definition
\begin{equation}\label{nu}
\Delta=\nu\frac{4 \pi H(r_0)^{1/2}}{ f'(r_0)} \, ,
\end{equation}
with $\nu=\pm 1$ to ensure that $\Delta>0$. 

From the QFT perspective, we would like to express the vacuum 
expectation values (VEVs) in terms of the sources, since this 
allows in principle 
for the calculation of higher $n$-point functions by derivation of
the one-point function at nonzero source. To proceed 
with this program, we note that the regularity condition on the 
gauge fields implies that
\begin{equation}
Q_i=h_i\mu_i \, .
\end{equation}
The existence of a region where $f(r_0)=0$ implies that 
\begin{equation}\label{q0}
{\tt q} =  -M r_0^{2} + \frac{h_{1} h_{2} h_{3}}{L^{2}} 
 \, .
\end{equation}
The mass parameter can be found replacing eq.\eqref{q0} into eq. \eqref{nu}, using eq.(\ref{fp-hi}). In fact,
\begin{equation}\label{eq:MASS}
M=\frac{h_{2}h_{3}+h_{1}h_{3}+h_{1}h_{2}}{L^{2}}
- 2\pi\nu\,\frac{\sqrt{h_{1}h_{2}h_{3}}}{\Delta}
 \, .
\end{equation}
The remaining equations are $Q_i^2-{\tt q}+Mq_i^2=0$ (which are equivalent to the Einstein equations). These  read 
\begin{align} \label{eq:Y1}
\mu_1^2=-\frac{h_{3}+h_{2}}{L^{2}}
+ 2\pi\nu\,\frac{\sqrt{h_{1}h_{2}h_{3}}}{h_{1}\,\Delta}
=\frac{h_{2}h_{3}}{h_{1}L^{2}}-\frac{M}{h_{1}}
 \, , \\ \label{eq:Y2}
 \mu_2^2=-\frac{h_{3}+h_{1}}{L^{2}}
+ 2\pi\nu\,\frac{\sqrt{h_{1}h_{2}h_{3}}}{h_{2}\,\Delta}
=\frac{h_{1}h_{3}}{h_{2}L^{2}}-\frac{M}{h_{2}}
  \, ,\\ \label{eq:Y3}
\mu_3^2=-\frac{h_{2}+h_{1}}{L^{2}}
+ 2\pi\nu\,\frac{\sqrt{h_{1}h_{2}h_{3}}}{h_{3}\,\Delta}
=\frac{h_{1}h_{2}}{h_{3}L^{2}}-\frac{M}{h_{3}}
 \, .
\end{align}
We readily see that $h_i> 0$ implies that $\nu=1$ and $\Delta< \infty$. The equations \eqref{eq:Y1}, \eqref{eq:Y2}, \eqref{eq:Y3} can be decoupled to obtain $h_i=h_i(\mu_1,\mu_2,\mu_3,\Delta)$. These yield two linear equations for (let us choose $h_1$ and $h_2$) and a quintic equation $P(Z)=0$ for the variable 
\begin{equation*}
 Z=\frac{\Delta^2}{4\pi^2 L^4} h_3.  
\end{equation*}
The quintic equation for $Z$ reads
\begin{equation} \label{eq:EOM_Z}
\begin{split}
P(Z) = &4 Z^{5}
+\left(8\psi_{2}^{2}-5+8\psi_{1}^{2}-4\psi_{3}^{2}\right) Z^{4}
+(14\psi_{1}^{2}\psi_{2}^{2}-2\psi_{3}^{2}+\psi_{3}^{4}
-6\psi_{2}^{2}\psi_{3}^{2}+5\psi_{1}^{4}+5\psi_{2}^{4}-6\psi_{3}^{2}\psi_{1}^{2}\\
&-6\psi_{1}^{2}+1-6\psi_{2}^{2}) Z^{3}
+\Big(-6\psi_{1}^{2}\psi_{2}^{2}-8\psi_{2}^{2}\psi_{1}^{2}\psi_{3}^{2}
-2\psi_{2}^{2}\psi_{3}^{2}-2\psi_{3}^{2}\psi_{1}^{2}
+\psi_{1}^{2}+\psi_{1}^{6}+\psi_{2}^{6}+\psi_{2}^{2}\\
&-2\psi_{2}^{4}\psi_{3}^{2}-2\psi_{2}^{4}
+\psi_{2}^{2}\psi_{3}^{4}+7\psi_{2}^{4}\psi_{1}^{2}
+\psi_{1}^{2}\psi_{3}^{4}+7\psi_{2}^{2}\psi_{1}^{4}
-2\psi_{1}^{4}-2\psi_{1}^{4}\psi_{3}^{2}\Big) Z^{2}
+\psi_{1}^{2}\psi_{2}^{2}
(-1-2\psi_{3}\\
&+\psi_{1}^{2}+\psi_{2}^{2}-\psi_{3}^{2})
\left(-1+2\psi_{3}+\psi_{1}^{2}+\psi_{2}^{2}-\psi_{3}^{2}\right) Z
-\psi_{2}^{4}\psi_{1}^{4}\, ,
\end{split}
\end{equation}
where we redefined 
\begin{equation}
  \mu_i=\frac{2\pi L}{\Delta}\psi_i.  
\end{equation} 
This completely determines the space of soliton solutions. Indeed, the polynomial  equation  $P(Z)=0$,  yields $Z=Z(\psi_{i})$, which can be used to find $\langle T_{tt} \rangle$. The insight that we can obtain from this analysis is that there can be several different solutions for a given value of the sources. Each solution corresponds to a different positive root of 
$P(Z)=0$. 

Each of these solutions correspond to a branch of moduli space, or 
a certain quantum phase, and will be dual to a corresponding 
branch of solutions in the gravity dual.

With this general formalism, it is satisfactory to recover known solutions. Let us do this.
\subsubsection{Recovering Previously Known Solutions}
By choosing different values  of $\psi_i$, we find previously known solutions. \\
\\
\textbf{Einstein-Maxwell-AdS.}
When $h_1=h_2=h_3$ the scalar fields vanish everywhere and the solution is the Einstein-Maxwell-AdS-Soliton of \cite{Anabalon:2021tua}. When $\psi_1=\psi_2=\psi_3 =\psi$ the quintic polynomial yields
\begin{equation}
P(Z)=(-1+Z)\,(Z+\psi^{2})^{2}\,(-Z+4Z^{2}+4Z\psi^{2}+\psi^{4})\;,
\end{equation}
which has two non-trivial solutions that exist provided $\psi^{2}<8^{-1}$. The root at $Z=1$ is unphysical as yields $\psi^2<1$ when replaced in eqs.\eqref{eq:Y1}, \eqref{eq:Y2}, \eqref{eq:Y3}.\\
\\
\textbf{Truncation to one scalar.} If $h_1=h_2$, then $\Phi_2=0$ and we recover the solution of \cite{Anabalon:2024che}. The quintic polynomial is now
\begin{equation}
P(Z)=(Z+\psi_{1}^{2})^{2}\,\bigl(
Z-5Z^{2}+4Z^{3}-2Z\psi_{1}^{2}+8Z^{2}\psi_{1}^{2}-\psi_{1}^{4}
+4Z\psi_{1}^{4}-2Z\psi_{2}^{2}-4Z^{2}\psi_{2}^{2}
-4Z\psi_{1}^{2}\psi_{2}^{2}+Z\psi_{2}^{4}
\bigr).
\end{equation}
To reproduce the cubic polynomial of equation (3.26) of \cite{Anabalon:2024che} it is necessary to find the equation 
satisfied by the combination $X=\frac{h_1}{h_3}$, in the case $\psi_2=\psi_1$. We find
\begin{equation}
0=\psi_{1}^{2} X^{3}
+\left(4\psi_{1}^{4}-4\psi_{3}^{2}\psi_{1}^{2}-\psi_{1}^{2}
+\psi_{3}^{4}-\psi_{3}^{2}\right) X^{2}
-\psi_{3}^{2}\left(4\psi_{1}^{2}-1-2\psi_{3}^{2}\right) X
+\psi_{3}^{4}\,,
\end{equation}
which is indeed in full agreement with 
\cite{Anabalon:2024che}.\\
\\
\textbf{Einstein-AdS.} The Horowitz-Myers solution has energy density
\begin{equation}
E_{0}=-\frac{\pi^4 L^3 }{2\kappa \Delta^4}\,.
\end{equation}
In our variables it corresponds to having no sources for the 
vectors, $\mu_i=0$. From eqs.\eqref{eq:Y1}, \eqref{eq:Y2}, 
\eqref{eq:Y3}, this implies that $h_ih_j =M L^2$ for $i\neq j$. This, substituted in eq.\eqref{eq:MASS}, yields $M=\frac{\pi^4L^6}{\Delta^4}$, and we find that $\langle T_{tt} \rangle=E_0$.

It is algebraically hard to find other generic solutions. This is ameliorated if we consider the massless limit $M=0$.

\subsubsection{The general massless solution}\label{susy-section}
\begin{figure}[ht]
  \centering
  \includegraphics[width=0.8\linewidth]{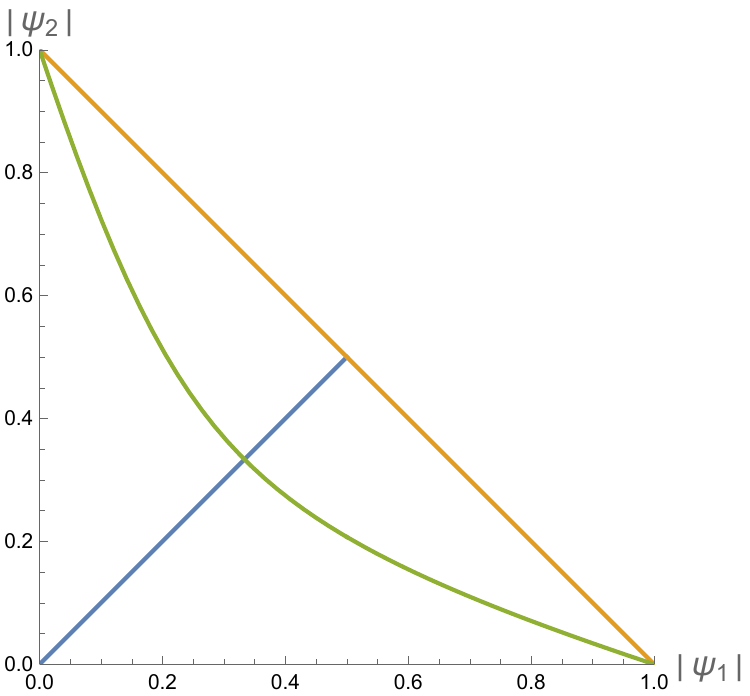}
  \caption{Moduli Space. Supersymmetric solutions exist below the orange line. Above the green line $\langle \mathcal{O}_1 \rangle<0$ and below the green line $\langle \mathcal{O}_1 \rangle>0$, $\langle \mathcal{O}_1 \rangle=0$ on the green line. The solutions that lie on the blue line have $\langle \mathcal{O}_2\rangle=0$ and are the supersymmetric solitons of \cite{Anabalon:2024che}. Above the blue line $\langle \mathcal{O}_2 \rangle>0$ and below the blue line $\langle \mathcal{O}_2\rangle<0$. The point of intersection between the green and blue lines happens at $\lvert \psi_1 \rvert=\lvert \psi_2\rvert=\lvert \psi_3\rvert=\frac{1}{3}$, which is the supersymmetric soliton of \cite{Anabalon:2021tua}.} 
  \label{fig:figure1}
\end{figure}

Dealing with the full quintic polynomial $P(Z)$ is very involved. We focus on 
the massless solutions of $P(Z)=0$. As we show 
in the next section, these are 
supersymmetric backgrounds. In this case, the zero energy condition 
$\langle T_{\mu \nu} \rangle=0$ implies that $M=0$ in 
eq.\eqref{eq:MASS}. We find that on this restricted {\em moduli 
space}
\begin{equation}
\sqrt{\frac{h_1h_2}{h_3}}+\sqrt{\frac{h_1h_3}{h_2}}+\sqrt{\frac{h_2h_3}{h_1}}=\frac{2\pi L^2}{\Delta}\, .
\end{equation}
It then follows from eqs.
\eqref{eq:Y1}, \eqref{eq:Y2}, \eqref{eq:Y3} that
\begin{equation}\label{eq:ContraintmuR}
\lvert \mu_1 \rvert+\lvert \mu_2 \rvert+\lvert \mu_3 \rvert=\frac{2\pi L}{\Delta}\, .
\end{equation}
The IR variables $h_i$ in terms of the sources $\mu_i$
are then given by
\begin{equation}
h_1 = L^2 \lvert \mu_2 \mu_3\rvert\, , \qquad h_2 = L^2 
\lvert \mu_1 \mu_3\rvert \, , \qquad h_3 = L^2 \lvert \mu_1 
\mu_2\rvert\, .
\end{equation}
The scalar VEVs are proportional to the following combinations
of the same sources,
\begin{equation}
q_1^2-q_2^2= h_1-h_2=L^2 \lvert \mu_3\rvert( \lvert\mu_2\rvert-\lvert\mu_1\rvert)\, ,
\end{equation}
\begin{equation}
q_1^2+q_2^2-2q_3^2=h_1+h_2-2h_3=L^2(\lvert \mu_2 \mu_3\rvert+\lvert \mu_1 \mu_3\rvert-2\lvert \mu_1 \mu_2\rvert).\label{qsmus}
\end{equation}
Expressing $\mu_i =(2\pi L/\Delta)\psi_i$, we find that
\begin{align}
\langle \mathcal{O}_1 \rangle&= \frac{q_1^2+q_2^2-2q_3^2}{2\kappa \sqrt{6} L} = \frac{(2 \pi)^2 L^3}{\kappa} \frac{1}{2\sqrt{6}\Delta^2} (\lvert \psi_{2}\psi_{3}\rvert+\lvert \psi_{1}\psi_{3}\rvert-2\lvert \psi_{1}\psi_{2}\rvert)
\, , \\
\langle \mathcal{O}_2 \rangle&=\frac{ q_1^2-q_2^2}{2\kappa \sqrt{2} L}= \frac{(2 \pi)^2 L^3 }{\kappa} \frac{1}{2\sqrt{2}\Delta^2} \lvert \psi_{3}\rvert\bigl(\lvert \psi_{2}\rvert-\lvert \psi_{1}\rvert\bigr)
\, ,\label{scalarexpf}
\end{align}
which are the strongly coupled, large-$N$, Yang-Mills VEVs in terms of the dimensionless variables $\psi_i$ (which are independent of the number of colors $N$, it only appears through the overall factor $\frac{ (2 \pi)^2 L^3}{\kappa}=N^2$). This vacuum state is characterized by the VEVS of the scalar as shown in Figure \ref{fig:figure1}.

The diagram of Figure \ref{fig:figure1} can be interpreted as the moduli space of supersymmetric solutions parameterized by the sources $(|\psi_1|,|\psi_2|)$. The lines and curves partition the space into regions characterized by different relative signs of the operator expectation values. Crossing a boundary does not lift the degeneracy of the vacuum but instead corresponds to a continuous change in the sign of a given VEV. All points in the diagram thus represent distinct supersymmetric vacua with identical vanishing energy at zero temperature and zero chemical potential.

This suggests another interpretation, as a phase diagram for quantum phases, 
with the blue and green lines (representing $\langle \mathcal{O}_2\rangle=0$ and 
$\langle  \mathcal{O}_1\rangle=0$, respectively)
separating qualitatively different quantum phases (defined by 
quantum fluctuations, not thermal fluctuations). The 
intersection is then a quantum critical point. 

Let us now study the Killing spinors preserved by these solutions with $M=0$.

\subsection{Supersymmetry}\label{susy-sectionN}

Let us label the scalar by $\Phi_{s}=(\Phi_{1},\Phi_{2})$. The Killing spinor equations for gauged STU model can be written in terms of the complex spinor $\epsilon$ defined in terms of the symplectic Majorana spinor $\tilde{\epsilon}^{a}$ as follows $\epsilon=\tilde{\epsilon}^{1}+\ri\tilde{\epsilon}^{2}$. The pair of spin 1/2 and the spin 3/2 variations are 
\begin{align}
\delta\lambda_{s} & =\sum_{i=1}^{3}\partial_{s}X_{i}\left[-\frac{1}{4X_{i}^{2}}\left(\not F^{i}+\ri\not{\dd}X_{i}\right)+\frac{\ri}{2L}\right]\epsilon=0\;,\label{varspin1/2}\\
\delta\psi_{\mu}\dd x^{\mu} & =\dd\epsilon+\frac{1}{4}\omega_{ab}\gamma^{ab}+\sum_{i=1}^{3}\left[\frac{\ri}{4!X_{i}}[2\not e\not F^{i}-6e^{a}(\iota_{a}F^{i})_{\slash}]+\frac{1}{3!L}\not eX_{i}-\frac{\ri}{2L}A^{i}\right]=0\;,
\end{align}
where we have defined the slash of a generic $p$-form $F_{p}=\frac{1}{p!}F_{a_{1}\dots a_{p}}e^{a_{1}}\wedge\dots\wedge e^{a_{p}}$ as
\begin{align}
\not F_{p} & \equiv(F_{p})_{/}\equiv\frac{1}{p!}F_{a_{1}\dots a_{p}}\gamma^{a_{1}\dots a_{p}}\,,
\end{align}
and $\iota_{a}$ stands for the contraction operator. In what follows, we choose the vielbein basis
    \begin{equation}
    \begin{aligned}
        e^{0} & =\frac{r}{L}H(r)^{1/6}\dd t\,,\qquad e^{1}=\frac{r}{L}H(r)^{1/6}\dd x\,,\qquad e^{2}=\frac{r}{L}H(r)^{1/6}\dd y\,,\\
        e^{3} & =\frac{H(r)^{1/6}}{f(r)^{1/2}}\dd r\,,\qquad e^{4}=\frac{\sqrt{f(r)}}{H(r)^{1/3}}\dd\varphi.
    \end{aligned}   
    \end{equation}

For the background (\ref{thebackground})-(\ref{A=000020Lambda}), the vanishing of the determinants of the spin 1/2 matrices in (\ref{varspin1/2}) requires that $M=0$. In this limit, the configuration is regular and there exist an $r_{0}$ such that $f(r_{0})=0$ when ${\tt q}>0$, hence
the configuration is still regular. The equations (\ref{varspin1/2})
for $s=1,2$ coincide and is given by
\begin{align}
[L\sqrt{\mathtt{q}}\gamma^{34}+\ri r^{2}Lf(r)^{1/2}\gamma^{3}+\ri r^{3}H(r)^{1/2}]\epsilon & =0\,.
\end{align}
This can be written as a projector equation
\begin{align}
 & \frac{1}{2}(1+\mathrm{x}(r)\gamma^{3}+\mathrm{y}(r)\gamma^{4})\epsilon=0\,,\label{proj=000020equation}
\end{align}
with
\begin{align}
\mathrm{x}(r)=\frac{r}{L}\frac{H(r)^{1/2}}{f(r)^{1/2}}\,,\qquad\mathrm{y}(r)=-\frac{\ri\sqrt{\mathtt{q}}}{r^{2}f(r)^{1/2}}\,,
\end{align}
satisfying $\mathrm{x}^{2}+\mathrm{y}^{2}=1,$ for any $r$. 

Regarding the spin 3/2 equation, in the coordinate basis, the component
$r$ reads
\begin{align}
\partial_{r}\epsilon+f(r)^{-1/2}\sum_{i}\frac{1}{H_{i}(r)}\left(\frac{\ri\sqrt{\mathtt{q}}}{3r^{3}}\gamma^{4}+\frac{H(r)^{1/2}}{6L}\gamma^{3}\right)\epsilon & =0\,.
\end{align}
By using the projector equation (\ref{proj=000020equation}), 
it can be recast in the form
\begin{align}
\partial_{r}\epsilon & =[\mathrm{a}(r)+\mathrm{b}(r)\gamma^{3}]\epsilon\,,\label{diff=000020eq=000020in=000020r=000020spinor}
\end{align}
with
\begin{align}
\mathrm{a}(r)=-\frac{1}{3r}\sum_{i}\frac{1}{H_{i}(r)}\,,\qquad\mathrm{b}(r)=-\frac{H(r)^{1/2}}{2Lf(r)^{1/2}}\sum_{i}\frac{1}{H_{i}(r)}\,.
\end{align}
Following \cite{Romans:1991nq}, the sufficient condition to find a common
solution of (\ref{proj=000020equation}) and (\ref{diff=000020eq=000020in=000020r=000020spinor})
is that
\begin{align}
\mathrm{x}^{\prime}(r)+2\mathrm{b}(r)\mathrm{y}(r)^{2} & =0\,,
\end{align}
which is satisfied in our case. Then, the spinor that satisfies (\ref{proj=000020equation})
and (\ref{diff=000020eq=000020in=000020r=000020spinor}) is given
by 
\begin{align}
\epsilon & =\frac{f(r)^{1/4}}{2H(r)^{1/6}}\left(\alpha_{+}(r)-\alpha_{-}(r)\gamma^{4}\right)(1-\gamma^{3})\chi(t,\varphi,x,y)\,,\\
\alpha_{\pm}(r) & \equiv\sqrt{1\pm\frac{rH(r)^{1/2}}{Lf(r)^{1/2}}}\,.
\end{align}
Given the projector equation (\ref{proj=000020equation}), it is straightforward
to show that the spin 3/2 equations along $t,\varphi,x,y$ reduce
to
\begin{align}
\partial_{t}\chi & =0\,,\nonumber \\
\partial_{\varphi}\chi+\frac{\ri\sqrt{\mathtt{q}}}{2L}\sum_{i}\mu_{i}\chi & =0\,,\\
\partial_{x}\chi & =0\,,\nonumber \\
\partial_{y}\chi & =0\,.\nonumber 
\end{align}
Therefore, the general solution to the Killing spinor equations is
\begin{align*}
\epsilon(r,\varphi) & =\exp\left(-\frac{\ri\varphi\sqrt{\mathtt{q}}}{2L}\sum_{i}\mu_{i}\right)\frac{f(r)^{1/4}}{2H(r)^{1/6}}\left(\alpha_{+}(r)-\alpha_{-}(r)\gamma^{4}\right)(1-\gamma^{3})\epsilon_0\,,
\end{align*}
\\
where $\epsilon_0$ is a constant complex spinor with four independent components. The projector in the solution reduces these to two, so the 5D background with $M=0$ preserves four supercharges.

Now that we have established that the solutions with $M=0$ preserve a fraction of the SUSY, we lift the solution to type IIB supergravity.

\section{Uplift to type IIB supergravity}\label{section-uplift}

The uplift to type IIB supergravity of the gauged STU model in $D=5$ was constructed in \cite{Cvetic:1999xp} as a compactification on a deformed $\Sp^{5}$. The 10-dimensional metric and five-form field strength are given by \footnote{We find the second term sign in $G_{5}$ is  opposite respect to \cite{Cvetic:1999xp}.  Our convention for the Hodge dual of a $p$-form
$F_{p}=\frac{1}{p!}F_{\mu_{1}\dots\mu_{p}}\dd x^{\mu_{1}}\wedge\dots\wedge\dd x^{\mu_{p}}$
is $\star F_{p}=\frac{\sqrt{-g}}{(D-p)!p!}F^{\mu_{1}\dots\mu_{p}}\epsilon_{\mu_{1}\dots\mu_{p}\nu_{1}\dots\nu_{D-p}}\dd x^{\nu_{1}}\wedge\dots\wedge\dd x^{\nu_{D-p}}$,
for a $D$-dimensional manifold with metric $g_{\mu\nu}$ and determinant
$g$ and $\epsilon_{12\dots D}=+1$. We have checked the equations of motion for this type IIB configuration. }
\begin{align*}
\dd s_{10}^{2} & =\tilde{\Delta}^{1/2}\dd s_{5}^{2}+L^{2}\tilde{\Delta}^{-1/2}\sum_{i=1}^{3}X_{i}^{-1}\left[\dd\nu_{i}^{2}+\nu_{i}^{2}\left(\dd\phi_{i}+\frac{1}{L}A_{i}\right)^{2}\right]\,,\\
F_{5} & =G_{5}+\star G_{5}\,,\\
G_{5} & =\frac{2}{L}\sum_{i}\left(X_{i}^{2}\nu_{i}^{2}-\tilde{\Delta}X_{i}\right)\star_{5}1
\mathbin{+}\frac{L}{2}\sum_{i}X_{i}^{-1}\star_{5}\dd X_{i}\wedge\dd(\nu_{i}^{2})+\\
 & \quad+L^{2}\sum_{i}X_{i}^{-2}\nu_{i}\dd\nu_{i}\wedge\left(\dd\phi_{i}+\frac{1}{L}A_{i}\right)\wedge\star_{5}F_{i}\,.
\end{align*}
The Hodge dual in 10-dimension
is denoted by $\star$, while the Hodge dual in five-dimensions with
respect to metric $\dd s_{5}^{2}$ is $\star_{5}$. We  have defined
\begin{equation}
\tilde{\Delta}  =\sum_{i}X_{i}\nu_{i}^{2}\,, ~~~~
\sum_{i}\nu_{i}^{2}  =1\,.\label{def-Deltat}
\end{equation}
Focusing on the metric, we define $\hat{f}(r)\equiv\frac{L^{2}}{r^{2}}f(r)$, so
the ten-dimensional metric reads 
\bea
\dd s_{10}^{2} & =&\tilde{\Delta}^{1/2}\frac{r^{2}}
{L^{2}}H(r)^{1/3}\left(\frac{\hat{f}(r)}
{H(r)}\dd\varphi^{2}+\frac{L^{4}}{r^{4}}\frac{\dd r^{2}}{\hat{f}
(r)}+\dd x_{1,2}^{2}\right)\cr
&&+L^{2}\tilde{\Delta}^{-1/2}\sum_{i=1}^{3}X_{i}^{-1}\left[\dd\nu_
{i}^{2}+\nu_{i}^{2}\left(\dd\phi_{i}+\frac{1}
{L}A_{i}\right)^{2}\right]\,,
\eea
Explicitly, the functions $X_{i}$ read
\begin{align}
X_{1}^{3}=\frac{H_{2}(r)H_{3}(r)}{H_{1}^{2}(r)}\,,\hspace*{0.6cm}X_{2}^{3}=\frac{H_{1}(r)H_{3}(r)}{H_{2}^{2}(r)}\,,\hspace*{0.6cm}X_{3}^{3}=\frac{H_{1}(r)H_{2}(r)}{H_{3}^{2}(r)}\,.
\end{align}
We consider the following parameterization for $\nu_{i}$ as
\begin{align}
\nu_{1}=\cos\theta\,\cos\psi\,,\qquad\nu_{2}=\cos\theta\,\sin\psi\,,\qquad\nu_{3}=\sin\theta\;.
\end{align}
With this, the metric reads
    \begin{align}
        \dd s_{10}^{2} & =\frac{r^{2}}{L^{2}}\tilde{\Delta}^{1/2}H(r)^{1/3}\left(\frac{\hat{f}(r)}{H(r)}\dd\varphi^{2}+\frac{L^{4}}{r^{4}}\frac{\dd r^{2}}{\hat{f}(r)}+\dd x_{1,2}^{2}\right)+\label{our10Dmetric}\\
        & \quad+\frac{L^{2}}{\tilde{\Delta}^{1/2}}\left\{ \mathtt{A}^{2}\dd\theta^{2}+\mathtt{B}^{2}\dd\psi^{2}+2\mathtt{C}\dd\theta\dd\psi+\sum_{i=1}^{3}\frac{\nu_{i}^{2}}{X_{i}}\left(\dd\phi_{i}+\frac{1}{L}A_{i}\right)^{2}\,\right\} \;.\nonumber 
    \end{align}
Here we have defined
\begin{align}
\mathtt{A}^{2} & =\sin^{2}\theta\left(\frac{\cos^{2}\psi}{X_{1}}+\frac{\sin^{2}\psi}{X_{2}}\right)+\frac{\cos^{2}\theta}{X_{3}}\,,~~
\mathtt{B}^{2}  =\cos^{2}\theta\left(\frac{\sin^{2}\psi}{X_{1}}+\frac{\cos^{2}\psi}{X_{2}}\right)\,,\label{def-Delta2}\\
\mathtt{C} & =\cos\theta\:\sin\theta\,\cos\psi\,\sin\psi\left(\frac{1}{X_{1}}-\frac{1}{X_{2}}\right)\,,
~~\tilde{\Delta} =\cos^{2}\theta(X_{1}\cos^{2}\psi+X_{2}\sin^{2}\psi)+X_{3}\sin^{2}\theta\,.\nonumber
\end{align}
These expressions simplify in the limit 
$q_{1}=q_{2}$,
for which $X_{1}=X_{2}$ and $\tilde{\Delta}$ is a function of 
$\theta$ and $r$. 
%
The Type IIB background is useful to explore the non-perturbative dynamics of ${\cal N}=4$ SYM deformed by VEVs and flowing to a gapped QFT in $(2+1)$ dimensions.  We briefly elaborate on the deformation  in the next section. One can also calculate interesting observables in the non-perturbative QFT, for example, Wilson and 't Hooft loops, entanglement entropy, masses of glueballs, etc. We leave this for future work \cite{inpreparation}.

\Comment{
\section{The deformation of ${\cal N}=4$ SYM dual to the 
AdS soliton solution and Q-balls}

\textcolor{red}{CN: let me write the things i do not understad of this section. After the section finishes, i will write (in blue colour) the way i think this section can be better written, using some of the material here. 
\begin{itemize}
    \item{i would not write as is written below that we make a 'relevant deformation'. This lead us to think that we are inserting an operator with dimension less than four. I think what we have are VEVs for certain operators of dimension two and dimension three}
    \item{I think the deformation by a relevant operator like proposed below $m\Phi_1^2$ (being $\Phi_1$ a chiral multiplet of $N=4$ SYM) is not allowed in supergravity, because these operators are known to acquire a very large anomalous dimension. In eq 4.1 below that deformation by massive chirals can not be reproduced in supergravity}
    \item{we should better explain what we mean with the Q-balls. In other words, our supergravity solution is static, the Q-balls are dynamical solutions. After the section finishes, i will write a subsection in blue colour, trying to suggest a way we could write things}
\end{itemize}}

The AdS solution corresponds in field theory 
to an ${\cal N}=1$ deformation of the ${\cal N}=4$ SYM theory 
 that breaks conformality in the IR, while keeping it in the 
 UV (so a relevant deformation).
It is a mass term deformation, giving extra 
terms in the superpotential. 
In the case of the solution with a single Abelian vector, 
the deformation term is $m_1\Phi_1$, which leads 
to a mass term for a complex scalar, $m_1^2 |Z|^2$, 
since the ${\cal N}=1$ chiral superfield is $\Phi_1=Z+...$. 
This breaks the global R-symmetry from $SO(6)$ 
to $SO(2)\times SO(4)\simeq U(1)\times SU(2)\times SU(2)$. 

But in the present case, the R-symmetry is broken from 
$SO(6)$ down to $SO(2)\times SO(2)\times SO(2)= 
U(1)\times U(1)\times U(1)$, corresponding to the 3 Abelian 
vectors in the supergravity solution. 
This necessarily means that we are adding 
3 mass terms to the 3 complex 
scalars $Z,W,V$ constructed from 
the $SO(6)$ scalars $X^I$, such that the 
scalar potential becomes
\be
{\cal V}={\rm Tr}\left[\sum_{I<J}[X^I,X^J]^2
+m_1^2 |Z|^2+m_2^2|W|^2+m_3^2|V|^2\right].
\ee
Here $m_1,m_2,m_3$ should be related to the 
holographic sources $\mu_i$
coming from the gauge field boundary values, 
$A_i=\mu_id\varphi$, most
likely as $m_i=\mu_i$.

But then, we must also consider the deformations due 
to the 2 scalars, $\Phi_1,\Phi_2$, with 
charges $(+1,+1,-2)$ and $(+1,-1,0)$ under the 
$U(1)\times U(1)\times U(1)$ groups, respectively. 
As the scalars are restrictions to these charges of the 
$\underline{20}'$ operator ${\rm Tr}\left[X^IX^J-\frac{1}{6}
\delta_{IJ}X^2\right]$, the corresponding scalar operators 
deforming the potential are 
\be
\Phi_1\rightarrow {\cal O}_1={\rm Tr}[Z^2+W^2-2V^2]\;,\;\;
\Phi_2\rightarrow {\cal O}_2={\rm Tr}[Z^2-W^2]\;,
\ee
and so in the effective action we must add to the potential 
${\cal V}$ a function of ${\cal O}_1\bar{\cal O}_1$ and 
${\cal O}_2\bar{\cal O}_2$, 
just like in \cite{Freedman:1999gk} it was argued that, in the 
generic case of a deformation by an operator corresponding 
to ${\cal O}_{20'}={\rm Tr}\left[X^IX^J-\frac{1}
{6}\delta_{IJ}X^2\right]$, the deformation was by 
\be
{\cal O}_{20'}^2+...=\left({\rm Tr}\left[X^IX^J-\frac{1}
{6}\delta_{IJ}X^2\right]\right)^2+...
\ee
(plus higher orders). 

The only remaining question is, what kind of terms 
and coefficients will they have? For that, we 
turn to the fact that the boundary $A^i\propto d\varphi$ 
sources were understood in \cite{Anabalon:2024che} as sources 
for Q-ball charge densities $J^\varphi_i$, 
in the 2+1-dimensional theory reduced on $\varphi$ (so that 
$\varphi$ is an internal direction). Indeed, 
then we could write Q-ball ans\"{a}tze for the complex scalars,
\be
Z(\varphi=\omega_1 t,\vec{x})=e^{i\omega_1t}Z(\vec{x})\;,\;\;\;
W(\varphi=\omega_2 t, \vec{x})=e^{i\omega_2 t}W(\vec{x})\;,\;\;\;
V(\varphi=\omega_3 t, \vec{x})=e^{i\omega_2 t}V(\vec{x})\;,
\ee
but, in order for these to generate actual Q-balls solutions 
\cite{Coleman:1985ki}\footnote{Note that a nonabelian version of 
Q-balls is possible \cite{Safian:1987pr}, though it is not needed 
here.}, we need the condition that $\omega_i<m_i$ and 
that the potential divided by the modulus squared of the scalar 
fields, ${\cal V}/|Z|^2, {\cal V}/|W|^2, {\cal V}/|V|^2$, 
starts off at a nonzero value at zero fields (since the scalar
masses need to be nonzero), 
and then has an absolute minimum at some nonzero value of 
the scalar 
VEVs. This is only possible if the next nonzero power 
in the potential, after the quadratic one, has a 
negative coefficient, while the one after that 
has a positive coefficient. 

Assuming also that only integer powers of ${\cal O}_1
\bar{\cal O}_1$ and ${\cal O}_2\bar{\cal O}_2$ appear,
this would mean that the terms to be added to ${\cal V}$
in the effective action are 
\be
-|\lambda_1| {\cal O}_1\bar{\cal O}_1 -|\lambda_2| {\cal 
O}_2\bar{\cal O}_2
+|\lambda'_1|\left({\cal O}_1\bar{\cal 
O}_1\right)^2+|\lambda'_2|\left({\cal O}_2\bar{\cal 
O}_2\right)^2+...
\ee
which then indeed leads to the 3 kinds of Q-balls, 
so to $J^\varphi_i\neq 0$. 
\subsection{What i think we might want to write if we can answer the questions in red}
\textcolor{blue}{
In this section we present some understanding of the dynamics that is taking place in the field theory dual to our background in Section \ref{section-uplift}.
\\
We are compactifying ${\cal N}=4$ SYM on a finite size circle, represented by the $\varphi$-coordinate in eq.(\ref{eq-2.15}). The boundary conditions for the fields are periodic for bosons and antiperiodic for fermions. This would break SUSY, were not for the presence of Wilson lines, represented in supergravity by the asymtotic values of the gauge fields $A^i=\mu_id\varphi$. These Wilson lines 'twist the compactification' (mixing R-symmetry with a part of the Lorentz group associated to translations in $\varphi$). This allows the existence of  massless fermions (partners of the gauge field). The total number of supercharges is four \textcolor{red}{CN: is this correct, right?}. Being this a field theory compactified on a twisted circle, it presents an infinite number of massive KK-modes. These come in SUSY multiplets. As we lower the energy and flow to the IR, these massive modes decouple and the theory  compactifies to $(2+1)$-dimensions, with Chern-Simons terms being generated in analogy with the study of \cite{Kumar:2024pcz, Cassani:2021fyv}.
\\
The asymptotic value of the fields, studied in Section \ref{sec:asymptotic} indicate that operators of dimension two (holographically related to the STU-model scalar fields $\Phi_1,\Phi_2$), currents of dimension three, associated with the supergravity fields $A^i_\mu$  and the operator $T_{\mu\nu}$ are acquiring VEVs. In the SUSY case, $M=0$ and $<T_{\mu\nu}>=0$.
\\
This deformation is a generalization of that studied in \cite{Freedman:1999gk}. In that paper the authors propose a {\it low energy effective action} written in terms of the six scalars $X_j$ and the composite field (chiral primary operator) ${\cal O}_{20'}=\text{Tr} \left( X_i X_j-\frac{\delta_{ij}}{6} (X_i)^2\right)$. The proposed effective field theory has a potential of the form $V\sim \left({\cal O}_{20'}\right)^2$.
\\
In the case of \cite{Freedman:1999gk}, there is no twisted compactification and  the supergravity solution is singular. In our work, we solve these singularities by smoothly ending the space at $r=r_0$ (before reaching the would be singularity), thanks to the presence of the fields $A^i$ and $\Phi_1, \Phi_2$ that are dual to VEVs of operators with  dimensions two and three. \\
\\
What follows is tentative and very heuristic in character, it must be read as a proposed effective dynamics at low energies.
In fact, we can propose an effective Lagrangian, written in terms of the composite fields holographically related to the scalars $\Phi_1, \Phi_2$.  These operators are,
\begin{eqnarray}
& & \Phi_1\rightarrow {\cal O}_1={\rm Tr}[Z^2+W^2-2V^2]\;,\;\;
\Phi_2\rightarrow {\cal O}_2={\rm Tr}[Z^2-W^2]\;.\\
& & \text{where we defined}~~Z=X_1+i X_2, ~W=X_3+i X_4,~~V=X_5+i X_6~\text{in}~{\cal N}=4~\text{SYM}\nonumber 
\end{eqnarray}
Notice that our compactification breaks $SO(6)\to U(1)^3$ and the operators ${\cal O}_1$ and ${\cal O}_2$ have charges (1,1,-2) and (1,-1,0) under $U(1)_1\times U(1)_2\times U(1)_3$. In this way the operators are inert and can be used in an effective description. 
\\
\\In the effective action, written in terms of the ${\cal O}_{1,2}$ operators, we must add to the potential 
${\cal V}$ a function of ${\cal O}_1\bar{\cal O}_1$ and 
${\cal O}_2\bar{\cal O}_2$, 
just like  the authors of \cite{Freedman:1999gk} did,
\\
\be
{\cal V}\sim -|\lambda_1| {\cal O}_1\bar{\cal O}_1 -|\lambda_2| {\cal 
O}_2\bar{\cal O}_2
+|\lambda'_1|\left({\cal O}_1\bar{\cal 
O}_1\right)^2+|\lambda'_2|\left({\cal O}_2\bar{\cal 
O}_2\right)^2+...
\ee
The presence of the gauge fields can be proposed to be associated with solitonic objects (Q-ball-like) in the low energy effective field theory.
\\
Indeed, we 
can turn to the fact that the boundary $A^i\propto \mu_i d\varphi$ 
sources were understood in \cite{Anabalon:2024che} as sources 
for Q-ball charge densities $J^\varphi_i$, 
in the 2+1-dimensional theory reduced on $\varphi$ (so that 
$\varphi$ is actually an internal, or R-symmetry, direction for the low energy effective theory). Then we could write Q-ball ans\"{a}tze for the complex scalars,
\be
Z(\varphi=\omega_1 t,\vec{x})=e^{i\omega_1t}Z(\vec{x})\;,\;\;\;
W(\varphi=\omega_2 t, \vec{x})=e^{i\omega_2 t}W(\vec{x})\;,\;\;\;
V(\varphi=\omega_3 t, \vec{x})=e^{i\omega_2 t}V(\vec{x})\;,
\ee
but, in order for these to generate actual Q-balls solutions 
\cite{Coleman:1985ki}\footnote{Note that a nonabelian version of 
Q-balls is possible \cite{Safian:1987pr}, though it is not needed 
here.}, we need the condition that $\omega_i<m_i$ and 
that the potential divided by the modulus squared of the scalar 
fields, ${\cal V}/|Z|^2, {\cal V}/|W|^2, {\cal V}/|V|^2$, 
starts off at a nonzero value at zero fields (since the scalar
masses need to be nonzero), 
and then has an absolute minimum at some nonzero value of 
the scalar 
VEVs. This is only possible if the next nonzero power 
in the potential, after the quadratic one, has a 
negative coefficient, while the one after that 
has a positive coefficient. 
\\
Assuming also that only integer powers of ${\cal O}_1
\bar{\cal O}_1$ and ${\cal O}_2\bar{\cal O}_2$ appear,
this would mean that the terms to be added to ${\cal V}$
in the effective potential are 
\be
{\cal V}_{eff}\sim -|\lambda_1| {\cal O}_1\bar{\cal O}_1 -|\lambda_2| {\cal 
O}_2\bar{\cal O}_2
+|\lambda'_1|\left({\cal O}_1\bar{\cal 
O}_1\right)^2+|\lambda'_2|\left({\cal O}_2\bar{\cal 
O}_2\right)^2+...
\ee
which then indeed leads to the 3 kinds of Q-balls, 
so to $J^\varphi_i\neq 0$. \textcolor{red}{CN: it would be good to say, given the dynamics of $Z,V,W$ what is the time evolutions of $O_1,O_2$. Any comment on the supergravity background being static and the Q-balls dynamical? }
}
\\
\\
\\
{\bf If we want to be even more conservative, i would write the following section}
\\
\\
\\
\\
}

{
\section{ Field Theory Interpretation of the Supersymmetric Branch}
In this section we interpret the massless ($M=0$) 
soliton solutions constructed above from the 
viewpoint of the dual field theory.
\\
We consider 4D $\mathcal{N}=4$ Super Yang-Mills theory 
on $\mathbb{R}^{1,2} \times \text{S}^1$, where the circle is 
parameterized by the coordinate $\varphi$ appearing 
in the boundary metric eq.(\ref{eq-2.15}). The 
compactification scale is set by the period $\Delta$ 
in eq.(\ref{periododefi}). In what follows, we also consider supersymmetry breaking spin-structure on the circle, that is, anti-periodic boundary conditions for the fermions. To restore 4D $\mathcal{N}=1$ worth of supersymmetry, in addition to the 
geometric compactification, we turn on background 
flat connections for the Cartan subgroup $\text{U(1)}^3 
\subset \text{SO(6)}_\text{R}$, thus breaking the global R-symmetry to\footnote{Since we restore 4D $\mathcal{N}=1$ SUSY, it is convenient to work in such conventions. In 4D $\mathcal{N}=1$ language, only the U(1)$_{\text{R}}\times$SU(3)$_{\text{F}}\subset$ SU(4) $\cong$ SO(6) subgroup of the R-symmetry is manifest. Our U(1)$^{3}$ background gauge fields are linear combinations of the U(1)$_{\text{R}}$ and U(1)$^{2}_{\text{F}}\subset$ SU(3)$_{\text{F}}$ symmetries, specifically 
\begin{equation}\label{eq:RFgaugefields5D}
    A_\RR = A_1 + A_2 + A_3, \qquad A_\FF = A_1 - A_2, \qquad A_{\FF'} = \frac{2}{3} \left( A_1 + A_2 - 2 A_3 \right). \\
\end{equation}} $\text{U(1)}_\text{R}\times \text{U(1)}^{2}_\text{F}$. In the bulk these correspond to the 
boundary values of the gauge fields,
\begin{equation}
A_i \;\longrightarrow\; \mu_i\, \dd \varphi .
\end{equation}

From the dual field theory point of view, a priori there is no constraint relating the $\mu_{i}$ (besides \eqref{eq:ContraintmuR}). On the other hand, the supergravity solution imposes an extra condition, due to eq.\eqref{eq:ParametersConstraint} and the regularity condition in eq. \eqref{eq:RegularityGaugeFields}.

The parameters $\mu_i$ are sources for the (conserved)
R-symmetry currents $J^\mu_i$. However, we note 
first that these are not quite normal sources, which 
would have been of the type $A_i\propto \dd t$ 
(the sources for Noether currents would normally 
have  electric-type charge), 
but these are of a new, 
different kind, similar to the case of a current 
in a neutral conducting wire: with overall current,
but no overall charge. We will come back to this 
fact later. 
The dimension of the R-current in this 
four-dimensional theory is $\Delta_J=3$, as in the 
standard case, and as dual to any (massless) gauge 
field in the bulk.

Since the theory is compactified 
on $\text{S}^1$, a (perhaps better) interpretation would be 
given in the reduced 2+1 dimensional field theory.
In particular, the dimension of a Noether current in 
(2+1) dimensions is two. Yet, now, this is of a new 
kind, since we obtain $\langle
J^a\rangle=0$ for $a=0,1,2$
(in {\it all} the field theory directions), and only 
$\langle J^\varphi\rangle\neq 0$, which is a 
``current component in an internal direction''.
%
\subsubsection*{Twisted Compactification and Supersymmetry}
Compactification of $\mathcal{N}=4$ SYM on a circle generically breaks supersymmetry (in our case, with periodic boundary conditions for bosons and anti-periodic for fermions, SUSY is broken). However, the presence of background R-symmetry Wilson lines implements a twisted compactification. The fermions pick up phases determined by their U(1)$_{\text{R}}\times$ U(1)$^{2}_{\text{F}}$ charges when transported around the circle. For special values of the Wilson lines, these phases compensate the SUSY-breaking effect of the compactification \cite{Kumar:2024pcz}. 

On the gravity side, the SUSY solutions arise precisely in the massless branch $M=0$. As shown in Section \ref{susy-section}, supersymmetry requires the relation

\begin{equation}
|\mu_1| + |\mu_2| + |\mu_3| = \frac{2\pi L}{\Delta}.
\end{equation}
This condition ensures the existence of Killing spinors in the bulk, selecting a supersymmetric vacuum of the compactified theory. Since the five-dimensional STU model has eight supercharges, the projector found in Section \ref{susy-sectionN} reduces this by one half, leaving four preserved supercharges. The dual infrared theory therefore preserves $\mathcal{N}=2$ supersymmetry in three dimensions.

At energies $E$ much less than $\frac{ 1}{\Delta}$, the Kaluza-Klein modes along the circle decouple, and the theory flows to an effectively $(2+1)$-dimensional supersymmetric theory characterized by the parameters $\mu_i$.
\subsubsection*{Scalar Operators and Vacuum Structure}
The asymptotic analysis of Section 
\ref{sec:asymptotic} shows that the bulk scalars 
$\Phi_1$ and $\Phi_2$ are dual to operators of dimension $\Delta_{{\cal O}_i}=2$  in the 
four-dimensional theory, and as such the only 
difference between sources and VEVs in the asymptotic
expansion ($\Delta=2\Rightarrow d-\Delta=2$)
is the presence or not of a log. But from 
eqs.(\ref{scalarexp}) we see that the scalars
admit only normalizable fall-offs: 
the logarithmic modes associated with sources 
are absent. Therefore, the dual operators acquire 
vacuum expectation values but are not explicitly 
sourced.

These operators belong to (are restrictions of)
the $\mathbf{20'}$
operator ${\rm Tr}\left[X^IX^J-\frac{1}{6}
\delta_{IJ}X^2\right]$ of SO(6)$_\RR$ 
and can be written in terms of the three complex scalars
charged under the three U(1) R-symmetries ({conversely under U(1)$_{\text{R}}\times$ U(1)$^{2}_{\text{F}}$ }), namely
\begin{equation}
Z = X_1 + i X_2, \qquad
W = X_3 + i X_4, \qquad
V = X_5 + i X_6.
\end{equation}
The operators in the QFT have VEVs proportional 
to $q_1^2+q_2^2-2q_3^2$ and $q_1^2-q_2^2$, 
implying a similar structure for the combination 
of scalars in ${\cal O}_1$ and ${\cal O}_2$, 
while the charges of the gauge fields
$A^i$ in the supersymmetric, $M=0$, gravity solution 
are $Q_i^2={\bf q}$. This implies that the 
operator VEVs should only respect a diagonal U(1) symmetry ({namely U(1)$_{\RR}$}). These operators ${\cal O}_1$, ${\cal O}_2$ are
\begin{equation}
\mathcal{O}_1 = \operatorname{Tr}\!\left(Z^2 + W^2 - 2V^2\right),
\qquad
\mathcal{O}_2 = \operatorname{Tr}\!\left(Z^2 - W^2\right).
\end{equation}
These operators have scaling dimension two at large 
$N$ and strong coupling, as seen from the near-AdS
expansion in eq.(\ref{scalarexp}), see also eq.(\ref{scalarexp2}). 
Their non-zero expectation values 
characterize different vacua of the compactified 
theory.
The compactification together with the Wilson lines breaks the global symmetry
\begin{equation}
\text{SO(6)}_\RR \;\longrightarrow\; \text{U(1)}^3,
\end{equation}
in agreement with the presence of three 
independent bulk gauge fields. 

Importantly, since the scalars are not sourced, the deformation of the theory is entirely due to the compactification and the background R-symmetry holonomies. The scalar VEVs arise dynamically and are fixed by the requirement of regularity in the bulk. In particular, on the supersymmetric branch $M=0$, the vacuum energy vanishes, $\langle T_{\mu\nu}\rangle = 0$, and the VEVs satisfy the relations derived in Section \ref{susy-section}.

Thus, the quintic structure uncovered in Section 2.4 encodes the space of allowed vacua compatible with a given choice of Wilson lines $\psi_i$,
\begin{equation}
\psi_i=\lim_{\rho \rightarrow \infty}\frac{1}{2\pi L}\oint_{S^1}A_i=\mu_i\frac{\Delta}{2\pi L} \, .
\end{equation}
Different positive roots correspond to distinct branches of the moduli space.


From the point of view of the reduced 2+1 dimensional 
field theory, the same story of scalars not being 
sourced is true: the dimension
of the ${\bf 20}'$ scalar operator is now 1 (it 
contains 
two scalars of dimension 1/2). As was 
found for  holographic superconductors
\cite{Gubser:2008px,Hartnoll:2008vx}, the case of 
$d=3, \Delta=1$ has no source, but rather its pair 
in the asymptotic expansion, with dimension
$d-\Delta=2$, just gives another kind of VEV (we have
two VEVs instead of one VEV and one source).
%
\subsubsection*{Heuristic effective infrared description}
It is natural to attempt to describe the low-energy dynamics in terms of the composite operators $\mathcal{O}_1$ and $\mathcal{O}_2$, in the spirit of effective descriptions proposed in related holographic contexts \cite{Freedman:1999gk, Anabalon:2024che}. The gravity analysis indicates that the vacuum structure is non-trivial and admits multiple branches, suggesting that the effective potential for these operators possesses several extrema.

While a derivation of the precise three-dimensional effective action lies beyond the scope of this work, the holographic results strongly constrain its structure.
The only explicit deformation parameters are the Wilson lines $\psi_i$. The scalar operators are not sourced but acquire VEVs determined dynamically. SUSY on the $M=0$ branch enforces the linear relation among the $\psi_i$
\begin{equation}
|\psi_1| + |\psi_2| + |\psi_3| = 1.
\end{equation}

These features indicate that the infrared theory should be viewed as a supersymmetric $(2+1)$-dimensional theory with background R-charge densities determined by $\psi_i$, and with vacuum expectation values for scalar bilinears in the $\mathbf{20'}$.

To gain a first insight into the structure of the 
low-energy (Wilsonian) effective action, we are inspired by
\cite{Freedman:1999gk}, who also analyzed 
holographic Coulomb branch deformations (although 
their gravity solutions were singular, due to the 
absence of the gauge fields). These authors argued that the 
lowest dimension terms in the effective action were 
the canonical kinetic term, the classical 
scalar potential ${\cal V}=\Tr\left[\sum_{I,J}
[X^I,X^J]^2\right]$ and, in the 
generic case of a deformation by an operator 
corresponding 
to ${\cal O}_{20'}={\rm Tr}\left[X^IX^J-\frac{1}
{6}\delta_{IJ}X^2\right]$, the deformation 
in the effective potential ${\cal V}_{\rm eff}$
was by 
\be
{\cal O}_{{\bf 20}'}^2+...=\left({\rm Tr}
\left[X^IX^J-\frac{1}
{6}\delta_{IJ}X^2\right]\right)^2+...
\ee
(plus higher orders). 

Applying this to our case, we would have a deformation
by $\bar {\cal O}_1{\cal O}_1+\bar {\cal O}_2{\cal
O}_2$. However, we also know that the effective 
potential must have a minimum for nonzero VEVs
$\langle {\cal O}_1\rangle\neq 0,\langle{\cal O}_2
\rangle\neq 0$ in the case of nonzero $q_i,\psi_i$, 
and that the minimum must be at zero potential,
because of supersymmetry.

That means that the effective potential must 
contain the following lowest order terms 
in $\bar{\cal O}_i{\cal O}_i$,
\be
{\cal V}_{\rm eff}=
m_1^2|{\cal O}_1|+m_2^2|{\cal O}_2|
-|\lambda_1| \bar{\cal O}_1{\cal O}_1 -|\lambda_2| 
\bar {\cal O}_2{\cal O}_2
+|\lambda'_1|\left(\bar{\cal O}_1{\cal 
O}_1\right)^2+|\lambda'_2|\left(\bar{\cal O}_2{\cal 
O}_2\right)^2+...\;,\label{Veff1}
\ee
where $\lambda_1,\lambda_2,\lambda'_1,\lambda'_2$
must be related to the dual deformation parameters
$q_i,\psi_i$. The terms with $\lambda_i,\lambda'_i$
are needed in order to have a minimum at a nonzero 
VEV for ${\cal O}_i$, and the terms with $m_i^2$ are 
needed in order for the minimum to be at  
${\cal V}_{\rm eff}=0$. 
For the relation between $\lambda_i,\lambda'_i$ and 
$q_i,\psi_i$, we can find the VEVs at the minimum, 
and then use  (\ref{qsmus}) and (\ref{scalarexpf})
to express them in terms of parameters. If $m_i\neq 
0$, we obtain a cubic equation, but if $m_i=0$, the 
relation becomes simply 
$(|\lambda_1|/(2|\lambda'_1|))^{1/2}=|\psi_2\psi_3|+
|\psi_1\psi_3|-2|\psi_1\psi_2|$ and 
$(|\lambda_
{2}|/(2|\lambda'_2|))^{1/2}=|\psi_3|(|\psi_2|
-|\psi_1|)$. 
We note that the above also respects 
the diagonal $U(1)$ symmetry, as necessary.
%
%
%
%
\\
\underline{\bf Interpretation in terms of Q-ball charge densities}
\\
At this point we remember that from the
point of view of a 2+1 dimensional field theory,
$J^\varphi$ was not a normal kind of current: its 
0,1,2 components vanish, and only the internal 
$\varphi$ component is nonzero. 
This is not a Noether current, 
and by the same token is also not a 
topological current. 
It could, however, be a 
current density for a {\em non-topological}
solution, which is not so constrained.

In 
\cite{Anabalon:2024che} it was proposed that the
explanation for this current is in terms of 
Q-ball charge densities. Indeed, consider a
Q-ball solution ansatz for the scalars that 
combines the fact that there is a $\varphi$ extra 
direction with the standard Q-ball ansatz of 
Coleman \cite{Coleman:1985ki}, 
\bea
&& Z(\varphi=\omega_1 t, \vec{x})
=e^{ik_1\varphi}Z(\vec{x})=e^{i\tilde \omega_1 t}Z(
\vec{x})\cr
&& W(\varphi=\omega_2 t, \vec{x})
=e^{ik_2\varphi}W(\vec{x})=e^{i\tilde \omega_2 t}W(
\vec{x})\cr
 && V(\varphi=\omega_3 t, \vec{x})
=e^{ik_3\varphi}V(\vec{x})=e^{i\tilde \omega_3 t}V(
\vec{x}).
\eea
where $\tilde \omega_i=k_i\omega_i$,
and consider the Q-ball current densities
\be
J^\varphi_i=i \left([(\d_\varphi \bar Z)Z-\d_\varphi Z \bar Z],[(\d_\varphi \bar W)W-\d_\varphi
W\bar W],[(\d_\varphi \bar V)V-\d_\varphi V \bar V]
\right)\;,
\ee
which are  nonzero.
Of course, {\em 
on the solutions}, 
we could also have $J_i^0\neq 0$. But we 
could arrange for the total electric charge 
$\int J_i^0$ of a distribution to be 
zero, yet the Q-ball densities to be nonzero. 
This is similar to the case of a neutral current wire,
when we can have a zero total charge, but nonzero 
current through the wire, by averaging over the 
positive static charges of the nuclei, and the 
negative moving charges of the electrons. Here, 
though, we have a new situation: both the overall 
charge and the overall current vanish, even though 
an individual Q-ball can have both, but only the 
``charge in the internal direction'', i.e, Q-ball 
charge, is nonzero, by averaging over Q-balls of
positive and negative electric charges, 
but same Q-ball charge.

Incidentally, this answers the question: how 
come the Q-ball ansatz is time-dependent, but the 
gravity solution is not? The point is
that {\em one} Q-ball gives a relation between 
coordinates, $\varphi=\omega t$, just like {\em one} 
pointlike electron moving gives $x=vt$. When averaging
over a continuum of solutions that dependence
goes away.

Then, in order  for the ansatz 
to generate actual Q-balls solutions, as Coleman 
showed
\cite{Coleman:1985ki}\footnote{Note that a nonabelian version of 
Q-balls is possible \cite{Safian:1987pr}, though it is not needed 
here.}, we need the condition that $\omega_i<m_i$
(for stability against decay), which means a nonzero
mass term, and 
that the potential divided by the modulus squared 
of the scalar 
fields, ${\cal V}_{\rm eff}/|Z|^2, {\cal V}_{\rm 
eff}/|W|^2, {\cal V}_{\rm eff}/|V|^2$, 
starts off at a nonzero value at zero fields (since 
the scalar masses need to be nonzero), 
and then has an absolute minimum at some 
nonzero value of 
the scalar VEVs. The simplest way to achieve this 
is to have a nonzero mass term, then the next power
of the fields to have a negative coefficient, 
and the following power to have a positive 
coefficient. Of course, in terms of {\em independent}
Q-ball charges, one should consider as fields not 
$Z,W,V$, but rather ${\cal O}_1,{\cal O}_2$, in which 
case the conditions are on ${\cal V}_{\rm eff}/|{\cal 
O}_1|, {\cal V}_{\rm eff}/|{\cal O}_2|$ instead.

But these are exactly the two conditions that we 
imposed on the effective potential in order to have 
a nonzero VEV and supersymmetry, leading to 
(\ref{Veff1})! 
So we are guaranteed to have the Q-ball solutions, 
and thus the currents $J_i^\varphi$, as we wanted.




}

\section{Conclusions and outlook}
In this work we constructed a new family of AdS soliton solutions of the
five-dimensional STU model with two independent scalar fields and three
Abelian gauge fields, and uplifted them to full ten-dimensional type IIB
supergravity. These backgrounds provide a holographic description of
$\mathcal N=4$ super Yang--Mills theory compactified on $\Sp^1$ in the presence
of three independent U(1)$^3\subset$ SO(6)$_\RR$ Wilson lines and with vacuum
expectation values for scalar bilinears in the ${\bf 20'}$ representation.
The resulting geometries are asymptotically AdS$_5$ and terminate smoothly
in the infrared, realizing a confining and gapped phase of the compactified
theory.

A central result of this paper is the explicit characterization of the space
of regular solutions. By imposing regularity in the interior together with
fixed boundary data for the Wilson lines and the compactification radius,
we showed that the bulk equations reduce to a quintic polynomial governing
the allowed solutions. Different positive roots correspond to distinct
branches of vacua of the dual field theory. In this way the gravitational
analysis provides a direct holographic realization of vacuum selection:
integration constants that would naively appear arbitrary are dynamically
fixed once the requirement of infrared regularity is imposed.

A particularly important sector arises in the massless limit $M=0$.
We demonstrated that these solutions preserve supersymmetry and admit a
transparent field theory interpretation. The supersymmetric branch is
characterized by a simple linear relation among the Wilson lines,
\begin{equation}
|\psi_1| + |\psi_2| + |\psi_3| = 1\;,
\end{equation}
which selects a moduli space of vacua with vanishing energy density.
From the dual point of view, the compactification and R-symmetry twists
induce vacuum expectation values for scalar operators while no explicit
sources for these operators are present. The resulting moduli space admits
a natural interpretation in terms of different quantum phases separated by
lines where the expectation values of composite operators change sign.
At the intersection of these loci the system exhibits a quantum critical
point, which in the gravitational description corresponds to a special
regular soliton solution.

The ten-dimensional uplift further clarifies the geometric origin of these
deformations. The Wilson lines in the gauge theory correspond to twists
along the angular directions of the internal S$^5$, while the scalar VEVs
appear as squashing modes of the compact space. This ten-dimensional
perspective provides a controlled framework to study non-perturbative
phenomena in the strongly coupled theory, including confinement and
symmetry breaking in the compactified $\mathcal N=4$ system.

Our results open several interesting directions for future investigation.
\begin{itemize}
\item{First, the geometries constructed here provide a natural laboratory to
compute non-perturbative observables in the confining phase of the
compactified theory. It would be particularly interesting to analyze
Wilson and 't~Hooft loops, glueball spectra, and entanglement entropy in
these backgrounds. Such observables should encode detailed information
about the vacuum structure and could help distinguish the different
branches of solutions identified in the gravitational analysis. We  do this in \cite{inpreparation}.}
\item{
Second, the quintic structure governing the solution space suggests the
existence of a rich phase diagram in the dual field theory. A systematic
study of the thermodynamics of these solutions, including finite
temperature generalizations and possible black hole counterparts, could
reveal phase transitions between different branches of vacua and clarify
the role of the supersymmetric locus as a critical boundary in parameter
space.}
\item{
Third, it would be interesting to understand the effective infrared
description of the compactified theory more directly from the field theory
side. Our holographic results strongly constrain the structure of the
low-energy effective potential for the scalar bilinears in the ${\bf 20'}$
multiplet. Deriving such an effective description explicitly would provide
a valuable bridge between the gravitational picture and the microscopic
dynamics of $\mathcal N=4$ SYM.}
\item{
Fourth: finding a nonlinear embedding of the five-dimensional solutions into
eleven dimensions, suggests a broader geometric
framework in which these backgrounds may arise. Exploring the M-theory
interpretation of these configurations, and understanding their relation to
brane constructions and higher-dimensional compactifications, could shed
light on the origin of the moduli space structure uncovered here.}
\end{itemize}
In summary, the new class of multi-charge AdS solitons constructed in this
paper provides a concrete holographic realization of vacuum structure and
quantum phases in compactified $\mathcal N=4$ super Yang--Mills theory.
We expect that further exploration of these geometries will yield new
insights into the interplay between supersymmetry, confinement, and
holography in strongly coupled gauge theories.

\section*{Acknowledgements}
We would like to thank Iosif Bena for discussions.
The work of HN is supported in part by  CNPq grant 
304583/2023-5 and FAPESP grant 2019/21281-4.
HN would also like to thank the ICTP-SAIFR for their support 
through FAPESP grant 2021/14335-0, and to CEA-Saclay for their 
hospitality during a part of this project. C. N. is supported by STFC’s grants ST/Y509644-1, ST/X000648/1 and ST/T000813/1. The work of AA is supported in part by the FONDECYT grants 1230853, 1242043, 1250133, 1262452 and 1262414 and by the FAPESP grant 2024/16864-9. The work of R.S. is supported by the Ram\'on y Cajal fellowship RYC2021-033794-I. R.S. also acknowledges support from grants from the Spanish government MCIU-22-PID2021-123021NB-I00, MCIU-25-PID2024-161500NB-I00 and principality of Asturias SV-PA-21-AYUD/2021/52177. R.S. thanks KU Leuven for the kind hospitality while some parts of this work were being completed. This work was supported by a short term scientific mission grant from the COST action CA22113 THEORY-CHALLENGES.

\bibliographystyle{JHEP.bst}
\bibliography{biblio.bib}

@article{Anabalon:2014fla,
    author = "Anabalon, Andres and Astefanesei, Dumitru and Martinez, Cristian",
    title = "{Mass of asymptotically anti{\textendash}de Sitter hairy spacetimes}",
    eprint = "1407.3296",
    archivePrefix = "arXiv",
    primaryClass = "hep-th",
    doi = "10.1103/PhysRevD.91.041501",
    journal = "Phys. Rev. D",
    volume = "91",
    number = "4",
    pages = "041501",
    year = "2015"
}

@article{Chatzis:2024kdu,
    author = "Chatzis, Dimitrios and Fatemiabhari, Ali and Nunez, Carlos and Weck, Peter",
    title = "{SCFT deformations via uplifted solitons}",
    eprint = "2406.01685",
    archivePrefix = "arXiv",
    primaryClass = "hep-th",
    doi = "10.1016/j.nuclphysb.2024.116659",
    journal = "Nucl. Phys. B",
    volume = "1006",
    pages = "116659",
    year = "2024"
}

@article{Cvetic:1999xp,
    author = "Cvetic, Mirjam and Duff, M. J. and Hoxha, P. and Liu, James T. and Lu, Hong and Lu, J. X. and Martinez-Acosta, R. and Pope, C. N. and Sati, H. and Tran, Tuan A.",
    title = "{Embedding AdS black holes in ten-dimensions and eleven-dimensions}",
    eprint = "hep-th/9903214",
    archivePrefix = "arXiv",
    reportNumber = "UPR-0840-T, CTP-TAMU-11-99, RU-99-4-B",
    doi = "10.1016/S0550-3213(99)00419-8",
    journal = "Nucl. Phys. B",
    volume = "558",
    pages = "96--126",
    year = "1999"
}

@article{Anabalon:2024che,
    author = "Anabal\'on, Andr\'es and Nastase, Horatiu and Oyarzo, Marcelo",
    title = "{Supersymmetric AdS solitons and the interconnection of different vacua of $ \mathcal{N} $ = 4 Super Yang-Mills}",
    eprint = "2402.18482",
    archivePrefix = "arXiv",
    primaryClass = "hep-th",
    doi = "10.1007/JHEP05(2024)217",
    journal = "JHEP",
    volume = "05",
    pages = "217",
    year = "2024"
}

@article{Behrndt:1998jd,
    author = "Behrndt, K. and Cvetic, Mirjam and Sabra, W. A.",
    title = "{Nonextreme black holes of five-dimensional N=2 AdS supergravity}",
    eprint = "hep-th/9810227",
    archivePrefix = "arXiv",
    reportNumber = "CAMS-98-6, HUB-EP-98-69, UPR-820-T",
    doi = "10.1016/S0550-3213(99)00243-6",
    journal = "Nucl. Phys. B",
    volume = "553",
    pages = "317--332",
    year = "1999"
}

@article{Anabalon:2025sok,
    author = "Anabal{\'o}n, Andr{\'e}s and Astefanesei, Dumitru and Oliva, Julio and Ortega, Gabriel and Urbina, Jorge",
    title = "{Phase Transitions and Black Hole Stability in Gauged N = 8 Supergravity}",
    eprint = "2512.05088",
    archivePrefix = "arXiv",
    primaryClass = "hep-th",
    month = "12",
    year = "2025"
}

@article{Anabalon:2024lgp,
    author = "Anabalon, Andres and Oliva, Julio",
    title = "{Plasma-Plasma Third Order Phase Transition from Type IIB Supergravity}",
    eprint = "2405.04611",
    archivePrefix = "arXiv",
    primaryClass = "hep-th",
    doi = "10.1103/PhysRevLett.133.121601",
    journal = "Phys. Rev. Lett.",
    volume = "133",
    number = "12",
    pages = "121601",
    year = "2024"
}

@article{Fatemiabhari:2025usn,
    author = "Fatemiabhari, Ali and Nastase, Horatiu and Nunez, Carlos and Roychowdhury, Dibakar",
    title = "{Holographic Krylov complexity in confining gauge theories}",
    eprint = "2511.22717",
    archivePrefix = "arXiv",
    primaryClass = "hep-th",
    month = "11",
    year = "2025"
}

@article{Chatzis:2025hek,
    author = "Chatzis, Dimitrios and Hammond, Madison and Itsios, Georgios and Nunez, Carlos and Zoakos, Dimitrios",
    title = "{Supersymmetric AdS Solitons, Coulomb Branch Flows and Twisted Compactifications}",
    eprint = "2511.18128",
    archivePrefix = "arXiv",
    primaryClass = "hep-th",
    reportNumber = "HU-EP-25/38",
    month = "11",
    year = "2025"
}

@article{Macpherson:2025pqi,
    author = "Macpherson, Niall T. and Merrikin, Paul and Nunez, Carlos and Stuardo, Ricardo",
    title = "{Twisted-circle compactifications of SQCD-like theories and holography}",
    eprint = "2506.15778",
    archivePrefix = "arXiv",
    primaryClass = "hep-th",
    doi = "10.1007/JHEP08(2025)146",
    journal = "JHEP",
    volume = "08",
    pages = "146",
    year = "2025"
}

@article{Nunez:2025gxq,
    author = "Nunez, Carlos and Roychowdhury, Dibakar",
    title = "{Timelike entanglement entropy: A top-down approach}",
    eprint = "2505.20388",
    archivePrefix = "arXiv",
    primaryClass = "hep-th",
    doi = "10.1103/vjyt-xc15",
    journal = "Phys. Rev. D",
    volume = "112",
    number = "2",
    pages = "026030",
    year = "2025"
}

@article{Nunez:2025ppd,
    author = "Nunez, Carlos and Roychowdhury, Dibakar",
    title = "{Interpolating between spacelike and timelike entanglement via holography}",
    eprint = "2507.17805",
    archivePrefix = "arXiv",
    primaryClass = "hep-th",
    doi = "10.1103/x3zd-llsx",
    journal = "Phys. Rev. D",
    volume = "112",
    number = "8",
    pages = "L081902",
    year = "2025"
}

@article{Fatemiabhari:2024lct,
    author = "Fatemiabhari, Ali and Nunez, Carlos and Piai, Maurizio and Rucinski, James",
    title = "{Stability of holographic confinement with magnetic fluxes}",
    eprint = "2411.16854",
    archivePrefix = "arXiv",
    primaryClass = "hep-th",
    doi = "10.1103/PhysRevD.111.066009",
    journal = "Phys. Rev. D",
    volume = "111",
    number = "6",
    pages = "066009",
    year = "2025"
}

@article{Kumar:2024pcz,
    author = "Kumar, S. Prem and Stuardo, Ricardo",
    title = "{Twisted circle compactification of $ \mathcal{N} $ = 4 SYM and its holographic dual}",
    eprint = "2405.03739",
    archivePrefix = "arXiv",
    primaryClass = "hep-th",
    doi = "10.1007/JHEP08(2024)089",
    journal = "JHEP",
    volume = "08",
    pages = "089",
    year = "2024"
}

@article{Barbosa:2024smw,
    author = "Barbosa, Marcelo and Nastase, Horatiu and Nunez, Carlos and Stuardo, Ricardo",
    title = "{Penrose limits of I-branes, twist-compactified D5-branes, and spin chains}",
    eprint = "2405.08767",
    archivePrefix = "arXiv",
    primaryClass = "hep-th",
    doi = "10.1103/PhysRevD.110.046015",
    journal = "Phys. Rev. D",
    volume = "110",
    number = "4",
    pages = "046015",
    year = "2024"
}

@article{Freedman:1999gk,
    author = "Freedman, D. Z. and Gubser, S. S. and Pilch, K. and Warner, N. P.",
    title = "{Continuous distributions of D3-branes and gauged supergravity}",
    eprint = "hep-th/9906194",
    archivePrefix = "arXiv",
    reportNumber = "CERN-TH-99-189, HUTP-99-A029, MIT-CTP-2877, USC-99-03",
    doi = "10.1088/1126-6708/2000/07/038",
    journal = "JHEP",
    volume = "07",
    pages = "038",
    year = "2000"
}

@article{Cvetic:2000zu,
    author = "Cvetic, Mirjam and Lu, Hong and Pope, C. N.",
    title = "{Consistent sphere reductions and universality of the Coulomb branch in the domain wall / QFT correspondence}",
    eprint = "hep-th/0004201",
    archivePrefix = "arXiv",
    reportNumber = "CTP-TAMU-11-00, UPR-884-T",
    doi = "10.1016/S0550-3213(00)00462-4",
    journal = "Nucl. Phys. B",
    volume = "590",
    pages = "213--232",
    year = "2000"
}

@article{Cvetic:1999xx,
    author = "Cvetic, Mirjam and Gubser, S. S. and Lu, Hong and Pope, C. N.",
    title = "{Symmetric potentials of gauged supergravities in diverse dimensions and Coulomb branch of gauge theories}",
    eprint = "hep-th/9909121",
    archivePrefix = "arXiv",
    reportNumber = "UPR-0856-T, CTP-TAMU-38-99, HUTP-99-A049, NSF-ITP-99-106",
    doi = "10.1103/PhysRevD.62.086003",
    journal = "Phys. Rev. D",
    volume = "62",
    pages = "086003",
    year = "2000"
}

@article{Witten:1998qj,
    author = "Witten, Edward",
    title = "{Anti de Sitter space and holography}",
    eprint = "hep-th/9802150",
    archivePrefix = "arXiv",
    reportNumber = "IASSNS-HEP-98-15",
    doi = "10.4310/ATMP.1998.v2.n2.a2",
    journal = "Adv. Theor. Math. Phys.",
    volume = "2",
    pages = "253--291",
    year = "1998"
}

@article{Anabalon:2021tua,
    author = "Anabalon, Andres and Ross, Simon F.",
    title = "{Supersymmetric solitons and a degeneracy of solutions in AdS/CFT}",
    eprint = "2104.14572",
    archivePrefix = "arXiv",
    primaryClass = "hep-th",
    doi = "10.1007/JHEP07(2021)015",
    journal = "JHEP",
    volume = "07",
    pages = "015",
    year = "2021"
}

@article{Fatemiabhari:2024aua,
    author = "Fatemiabhari, Ali and Nunez, Carlos",
    title = "{From conformal to confining field theories using holography}",
    eprint = "2401.04158",
    archivePrefix = "arXiv",
    primaryClass = "hep-th",
    doi = "10.1007/JHEP03(2024)160",
    journal = "JHEP",
    volume = "03",
    pages = "160",
    year = "2024"
}

@article{Chatzis:2025dnu,
    author = "Chatzis, Dimitrios and Hammond, Madison and Itsios, Georgios and Nunez, Carlos and Zoakos, Dimitrios",
    title = "{Universal Observables, SUSY RG-Flows and Holography}",
    eprint = "2506.10062",
    archivePrefix = "arXiv",
    primaryClass = "hep-th",
    month = "6",
    year = "2025"
}

@article{Maldacena:1997re,
    author = "Maldacena, Juan Martin",
    title = "{The Large $N$ limit of superconformal field theories and supergravity}",
    eprint = "hep-th/9711200",
    archivePrefix = "arXiv",
    reportNumber = "HUTP-97-A097, HUTP-98-A097",
    doi = "10.4310/ATMP.1998.v2.n2.a1",
    journal = "Adv. Theor. Math. Phys.",
    volume = "2",
    pages = "231--252",
    year = "1998"
}

@article{Papadimitriou:2004ap,
    author = "Papadimitriou, Ioannis and Skenderis, Kostas",
    editor = "Biquard, O.",
    title = "{AdS / CFT correspondence and geometry}",
    eprint = "hep-th/0404176",
    archivePrefix = "arXiv",
    reportNumber = "ITFA-2004-17",
    doi = "10.4171/013-1/4",
    journal = "IRMA Lect. Math. Theor. Phys.",
    volume = "8",
    pages = "73--101",
    year = "2005"
}

@article{Castellani:2024ial,
    author = "Castellani, Federico and Nunez, Carlos",
    title = "{Holography for confined and deformed theories: TsT-generated solutions in type IIB supergravity}",
    eprint = "2410.00094",
    archivePrefix = "arXiv",
    primaryClass = "hep-th",
    doi = "10.1007/JHEP12(2024)155",
    journal = "JHEP",
    volume = "12",
    pages = "155",
    year = "2024"
}

@article{Chatzis:2024top,
    author = "Chatzis, Dimitrios and Fatemiabhari, Ali and Nunez, Carlos and Weck, Peter",
    title = "{Conformal to confining SQFTs from holography}",
    eprint = "2405.05563",
    archivePrefix = "arXiv",
    primaryClass = "hep-th",
    doi = "10.1007/JHEP08(2024)041",
    journal = "JHEP",
    volume = "08",
    pages = "041",
    year = "2024"
}

@article{Horowitz:1998ha,
    author = "Horowitz, Gary T. and Myers, Robert C.",
    title = "{The AdS / CFT correspondence and a new positive energy conjecture for general relativity}",
    eprint = "hep-th/9808079",
    archivePrefix = "arXiv",
    reportNumber = "NSF-ITP-98-076, MCGILL-98-13",
    doi = "10.1103/PhysRevD.59.026005",
    journal = "Phys. Rev. D",
    volume = "59",
    pages = "026005",
    year = "1998"
}

@article{Macpherson:2024qfi,
    author = "Macpherson, Niall T. and Merrikin, Paul and Stuardo, Ricardo",
    title = "{Circle compactifications of Minkowski$_{D}$ solutions, flux vacua and solitonic branes}",
    eprint = "2412.15102",
    archivePrefix = "arXiv",
    primaryClass = "hep-th",
    doi = "10.1007/JHEP08(2025)143",
    journal = "JHEP",
    volume = "08",
    pages = "143",
    year = "2025"
}

@article{Giliberti:2024eii,
    author = "Giliberti, Mauro and Fatemiabhari, Ali and Nunez, Carlos",
    title = "{Confinement and screening via holographic Wilson loops}",
    eprint = "2409.04539",
    archivePrefix = "arXiv",
    primaryClass = "hep-th",
    doi = "10.1007/JHEP11(2024)068",
    journal = "JHEP",
    volume = "11",
    pages = "068",
    year = "2024"
}

@article{Jokela:2025cyz,
    author = "Jokela, Niko and Kastikainen, Jani and Nunez, Carlos and Pen{\'\i}n, Jos{\'e} Manuel and Ruotsalainen, Helime and Subils, Javier G.",
    title = "{On entanglement c-functions in confining gauge field theories}",
    eprint = "2505.14397",
    archivePrefix = "arXiv",
    primaryClass = "hep-th",
    reportNumber = "HIP-2025-16/TH",
    month = "5",
    year = "2025"
}

@article{Gubser:2008px,
    author = "Gubser, Steven S.",
    title = "{Breaking an Abelian gauge symmetry near a black hole horizon}",
    eprint = "0801.2977",
    archivePrefix = "arXiv",
    primaryClass = "hep-th",
    reportNumber = "PUPT-2255",
    doi = "10.1103/PhysRevD.78.065034",
    journal = "Phys. Rev. D",
    volume = "78",
    pages = "065034",
    year = "2008"
}

@article{Romans:1991nq,
    author = "Romans, L. J.",
    title = "{Supersymmetric, cold and lukewarm black holes in cosmological Einstein-Maxwell theory}",
    eprint = "hep-th/9203018",
    archivePrefix = "arXiv",
    reportNumber = "PRINT-92-0114 (JPL,CAL-TECH)",
    doi = "10.1016/0550-3213(92)90684-4",
    journal = "Nucl. Phys. B",
    volume = "383",
    pages = "395--415",
    year = "1992"
}

@article{Fatemiabhari:2026goj,
    author = "Fatemiabhari, Ali and Nunez, Carlos",
    title = "{Krylov Complexity, Confinement and Universality}",
    eprint = "2602.17757",
    archivePrefix = "arXiv",
    primaryClass = "hep-th",
    month = "2",
    year = "2026"
}

@article{inpreparation,
    author = Present-authors,
    title = "{Present authors. In preparation}"
}

@article{Nunez:2023nnl,
    author = "Nunez, Carlos and Oyarzo, Marcelo and Stuardo, Ricardo",
    title = "{Confinement in (1 + 1) dimensions: a holographic perspective from I-branes}",
    eprint = "2307.04783",
    archivePrefix = "arXiv",
    primaryClass = "hep-th",
    doi = "10.1007/JHEP09(2023)201",
    journal = "JHEP",
    volume = "09",
    pages = "201",
    year = "2023"
}

@article{Nunez:2023xgl,
    author = "Nunez, Carlos and Oyarzo, Marcelo and Stuardo, Ricardo",
    title = "{Confinement and D5 branes}",
    eprint = "2311.17998",
    archivePrefix = "arXiv",
    primaryClass = "hep-th",
    month = "11",
    year = "2023"
}

@article{Bianchi:2001kw,
    author = "Bianchi, Massimo and Freedman, Daniel Z. and Skenderis, Kostas",
    title = "{Holographic renormalization}",
    eprint = "hep-th/0112119",
    archivePrefix = "arXiv",
    reportNumber = "MIT-CTP-3166, PUTP-1999, DAMTP-2001-63, ROM2F-2001-30",
    doi = "10.1016/S0550-3213(02)00179-7",
    journal = "Nucl. Phys. B",
    volume = "631",
    pages = "159--194",
    year = "2002"
}

@article{Balasubramanian:1999re,
    author = "Balasubramanian, Vijay and Kraus, Per",
    title = "{A Stress tensor for Anti-de Sitter gravity}",
    eprint = "hep-th/9902121",
    archivePrefix = "arXiv",
    reportNumber = "HUTP-99-A002, EFI-99-6, NSF-ITP-98-132",
    doi = "10.1007/s002200050764",
    journal = "Commun. Math. Phys.",
    volume = "208",
    pages = "413--428",
    year = "1999"
}

@article{Skenderis:2002wp,
    author = "Skenderis, Kostas",
    editor = "de Wit, B. and Vandoren, S.",
    title = "{Lecture notes on holographic renormalization}",
    eprint = "hep-th/0209067",
    archivePrefix = "arXiv",
    reportNumber = "PUTP-2047",
    doi = "10.1088/0264-9381/19/22/306",
    journal = "Class. Quant. Grav.",
    volume = "19",
    pages = "5849--5876",
    year = "2002"
}

@article{Coleman:1985ki,
    author = "Coleman, Sidney R.",
    title = "{Q-balls}",
    reportNumber = "HUTP-85/A050",
    doi = "10.1016/0550-3213(86)90520-1",
    journal = "Nucl. Phys. B",
    volume = "262",
    number = "2",
    pages = "263",
    year = "1985",
    note = "[Addendum: Nucl.Phys.B 269, 744 (1986)]"
}

@article{Safian:1987pr,
    author = "Safian, Alexander M. and Coleman, Sidney R. and Axenides, Minos",
    title = "{SOME NONABELIAN Q BALLS}",
    reportNumber = "HUTP-87/A048, DOE-ER-40048-09-P7",
    doi = "10.1016/0550-3213(88)90315-X",
    journal = "Nucl. Phys. B",
    volume = "297",
    pages = "498--514",
    year = "1988"
}

@article{Cassani:2021fyv,
    author = "Cassani, Davide and Komargodski, Zohar",
    title = "{EFT and the SUSY Index on the 2nd Sheet}",
    eprint = "2104.01464",
    archivePrefix = "arXiv",
    primaryClass = "hep-th",
    doi = "10.21468/SciPostPhys.11.1.004",
    journal = "SciPost Phys.",
    volume = "11",
    pages = "004",
    year = "2021"
}

@article{Hartnoll:2008vx,
    author = "Hartnoll, Sean A. and Herzog, Christopher P. and Horowitz, Gary T.",
    title = "{Building a Holographic Superconductor}",
    eprint = "0803.3295",
    archivePrefix = "arXiv",
    primaryClass = "hep-th",
    reportNumber = "NSF-KITP-08-38, PUPT-2261",
    doi = "10.1103/PhysRevLett.101.031601",
    journal = "Phys. Rev. Lett.",
    volume = "101",
    pages = "031601",
    year = "2008"
}

\end{document}